\documentclass[twocolumn,tighten]{aastex62}

\usepackage{lipsum}  
\usepackage{soul}
\newcommand\SFRValue{1.71\pm0.21} %log
\newcommand\StelMassValue{9.00\pm0.32} %log
\newcommand\MeanMagValue{12\pm3} 
\newcommand\mDotValue{1.73\pm0.23} %log
\newcommand\etaValue{0.04\pm0.34} %log
\newcommand\voutValue{-144\pm79} %v_out
\newcommand\colDenValue{20.81\pm0.08} %\log(N(HI))
\newcommand\colDenCovValue{20.56\pm0.05} %\log(N(HI)f_{cov})
\newcommand{\fcovValue}{0.56\pm0.06}
\usepackage{float}
\newcommand\cfValue{0.4\pm0.2} 
\newcommand\RValue{$3.5$ kpc} %
\newcommand\deltaRValue{$2$ kpc}

\newcommand\srcRedshiftErr{1.86569 \pm 0.00031} \newcommand\srcRedshiftNebErr{1.865947 \pm 0.000031} 
\newcommand\srcRedshiftNeb{1.865947} 

    \usepackage{placeins}

\newcommand{\HI}{H\tiny{ }\footnotesize{I}\normalsize{ }}
\newcommand{\HII}{H\tiny{ }\footnotesize{II}\normalsize{ }}

\usepackage{amsmath}
\newcommand{\kms}{\ifmmode\,{\rm km}\,{\rm s}^{-1}\else km$\,$s$^{-1}$\fi}
\newcommand{\Lya}{\ifmmode\,{\rm Ly}{\rm \alpha}\else Ly$\alpha$\fi}
\newcommand{\Msun}{\mathrm{M}_{\sun}}

\usepackage{lipsum}  
\defcitealias{Dopita16}{D16}

\usepackage{lineno}
\newcommand{\allcaps}[1]{\verb!#1!}

\graphicspath{{./}{}}
\shorttitle{Spatially resolved outflows in CSWA13}
\shortauthors{Vasan G.C. et al.}

\begin{document}
\title{Spatially Resolved Galactic Winds at Cosmic Noon: Outflow Kinematics and Mass Loading in a Lensed Star-Forming Galaxy at $z=1.87$}

\correspondingauthor{Keerthi Vasan G.C.}
\email{kvch@ucdavis.edu}

\author[0000-0002-2645-679X]{Keerthi Vasan G.C.}
\affiliation{Department of Physics and Astronomy, University of California, Davis, 1 Shields Avenue, Davis, CA 95616, USA}
\affiliation{The Observatories of the Carnegie Institution for Science, 813 Santa Barbara Street, Pasadena, CA 91101, USA}

\author[0000-0001-5860-3419]{Tucker Jones}
\affiliation{Department of Physics and Astronomy, University of California, Davis, 1 Shields Avenue, Davis, CA 95616, USA}

\author[0000-0002-5558-888X]{Anowar J. Shajib}
\affiliation{Department of Astronomy \& Astrophysics, The University of Chicago, Chicago, IL 60637, USA}
\affiliation{Kavli Institute for Cosmological Physics, University of Chicago, Chicago, IL 60637, USA}
\affiliation{NHFP Einstein Fellow}

\author{Sunny Rhoades}
\affiliation{Department of Physics and Astronomy, University of California, Davis, 1 Shields Avenue, Davis, CA 95616, USA}

\author{Yuguang Chen}
\affiliation{Department of Physics and Astronomy, University of California, Davis, 1 Shields Avenue, Davis, CA 95616, USA}

\author[0000-0003-4792-9119]{Ryan L. Sanders}\affiliation{Department of Physics and Astronomy, University of Kentucky, 505 Rose Street, Lexington, KY 40506, USA}

\author{Daniel P. Stark}
\affiliation{Steward Observatory, University of Arizona, 933 N Cherry Ave, Tucson, AZ 85721, USA}

\author{Richard S. Ellis}
\affiliation{Department of Physics and Astronomy, University College London, Gower Street, London WC1E 6BT, UK}

\author[0000-0003-4570-3159]{Nicha Leethochawalit}
\affiliation{National Astronomical Research Institute of Thailand (NARIT), Mae Rim, Chiang Mai, 50180, Thailand}

\author[0000-0003-1362-9302]{Glenn G. Kacprzak}
\affiliation{Centre for Astrophysics and Supercomputing, Swinburne University of Technology, Hawthorn, Victoria 3122, Australia}
\affiliation{ARC Centre of Excellence for All Sky Astrophysics in 3 Dimensions (ASTRO 3D), Australia}

\author[0000-0002-2784-564X]{ Tania M. Barone }
\affiliation{Centre for Astrophysics and Supercomputing, Swinburne University of Technology, Hawthorn, Victoria 3122, Australia}
\affiliation{ARC Centre of Excellence for All Sky Astrophysics in 3 Dimensions (ASTRO 3D), Australia}
\affiliation{School of Physics, University of New South Wales, Kensington, Australia}

\author[0000-0002-3254-9044]{Karl Glazebrook }
\affiliation{Centre for Astrophysics and Supercomputing, Swinburne University of Technology, Hawthorn, Victoria 3122, Australia}
\affiliation{ARC Centre of Excellence for All Sky Astrophysics in 3 Dimensions (ASTRO 3D), Australia}

\author[0000-0001-9208-2143]{Kim-Vy H. Tran}
\affiliation{School of Physics, University of New South Wales, Kensington, Australia}
\affiliation{ARC Centre of Excellence for All Sky Astrophysics in 3 Dimensions (ASTRO 3D), Australia}

\author{Hannah Skobe}
\affiliation{Department of Astronomy \& Astrophysics, The University of Chicago, Chicago, IL 60637, USA}

\author{Kris Mortensen}
\affiliation{Department of Physics and Astronomy, University of California, Davis, 1 Shields Avenue, Davis, CA 95616, USA}

\author{Ivana Barisic}
\affiliation{Department of Physics and Astronomy, University of California, Davis, 1 Shields Avenue, Davis, CA 95616, USA}

\begin{abstract}
We study the spatially resolved outflow properties of CSWA13, an intermediate mass ($M_*=10^{9}~\Msun$), gravitationally lensed star-forming galaxy at $z=1.87$. 
We use Keck/KCWI to map outflows in multiple rest-frame ultraviolet ISM absorption lines, along with fluorescent \ion{Si}{2}* emission, and nebular emission from \ion{C}{3}] tracing the local systemic velocity. 
The spatial structure of outflow velocity mirrors that of the nebular kinematics, which we interpret to be a signature of a young galactic wind that is pressurizing the ISM of the galaxy but is yet to burst out. From the radial extent of \ion{Si}{2}* emission, we estimate that the outflow is largely encapsulated within \RValue. We explore the geometry (e.g., patchiness) of the outflow by measuring the covering fraction at different velocities, finding that the maximum covering fraction is at velocities $v\simeq-150$~\kms. 
Using the outflow velocity ($v_{out}$), radius ($R$), column density ($N$), and solid angle ($\Omega$) based on the covering fraction, we measure the mass loss rate $\log\dot{m}_{out}/(\Msun\text{yr}^{-1}) = \mDotValue$ and mass loading factor $\log\eta = \etaValue$ for the low-ionization outflowing gas in this galaxy. These values are relatively large and the bulk of the outflowing gas is moving with speeds less than the escape velocity of the galaxy halo, suggesting that the majority of outflowing mass will remain in the circumgalactic medium and/or recycle back into the galaxy. The results support a picture of high outflow rates transporting mass and metals into the inner circumgalactic medium, providing the gas reservoir for future star formation. 

\end{abstract}

% Unified Astronomy Thesaurus concepts: 

\section{Introduction}

\begin{figure*}
    \begin{minipage}{0.6\linewidth}
       \includegraphics[width=0.98\linewidth]{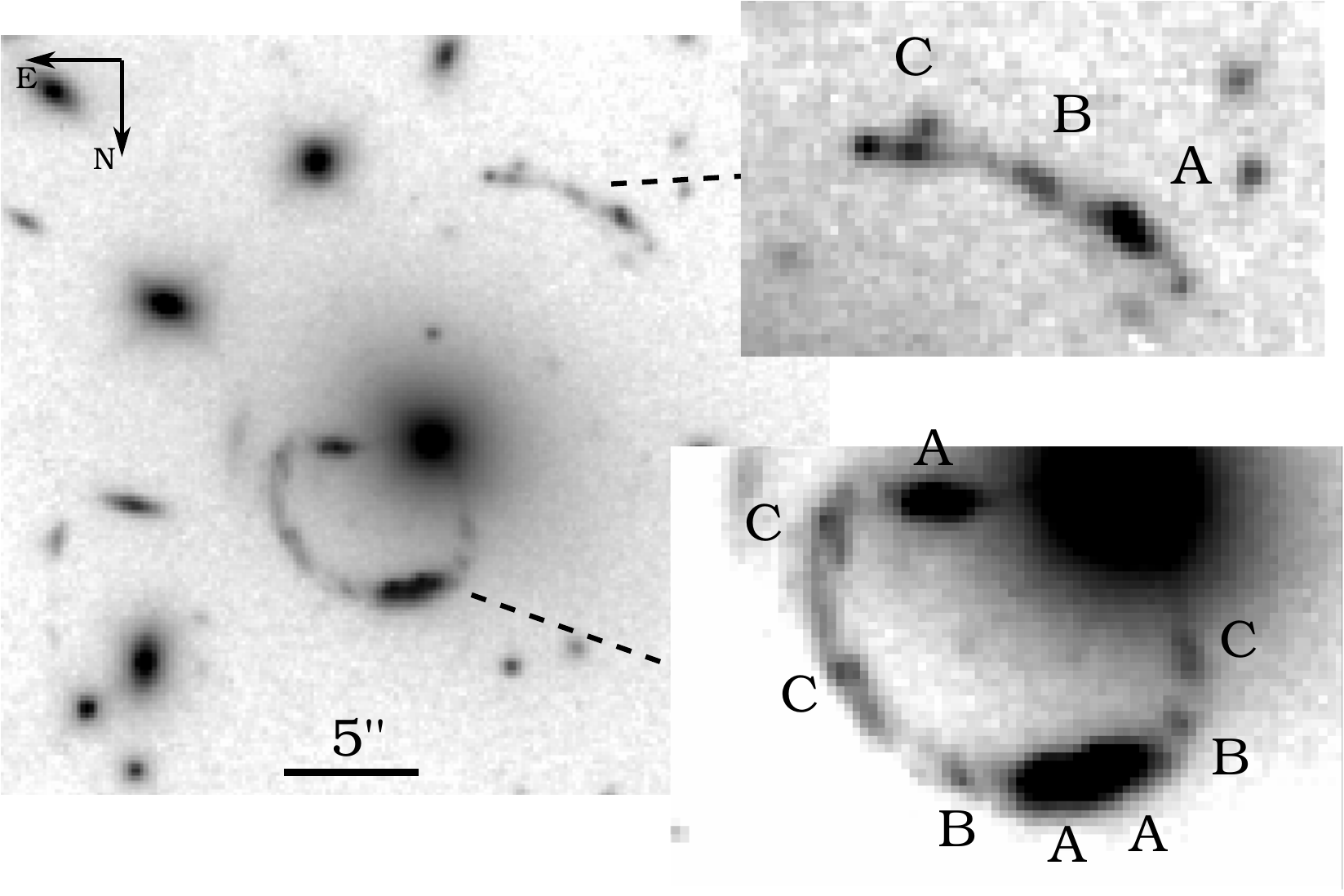} 
    \end{minipage}
    \begin{minipage}{0.4\linewidth}
        \centering
        \Large{KCWI color image}
       \includegraphics[width=0.8\linewidth]{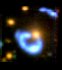}
    \end{minipage}    
    \caption{\emph{Left:} HST near-infrared image of CSWA13 in the F140W filter, which probes rest-frame optical wavelengths (B/g-band) at $z\simeq1.87$. East is left and North is down. The multiple images of the three bright distinct star-forming clumps of the galaxy are denoted as A, B and C. 
    Identification of these multiply imaged regions is also confirmed by the KCWI spectra, which show distinct velocity profiles for each source-plane region (see Section~\ref{subsec:neb-kinematics}). \emph{Right:} KCWI color image centered on the main arc and deflector. The color channels were constructed by summing three broad wavelength regions (B: 3530--3930~\AA, G: 4230--4630~\AA, R: 4930--5330~\AA) of the datacube. The lensed galaxy images form prominent bright blue arcs.}

    \label{fig:CSWA13-system-intro}
\end{figure*}
\begin{figure*}[!ht]
    \centering
    \Large{(a)}\\
    \includegraphics[width=\linewidth]{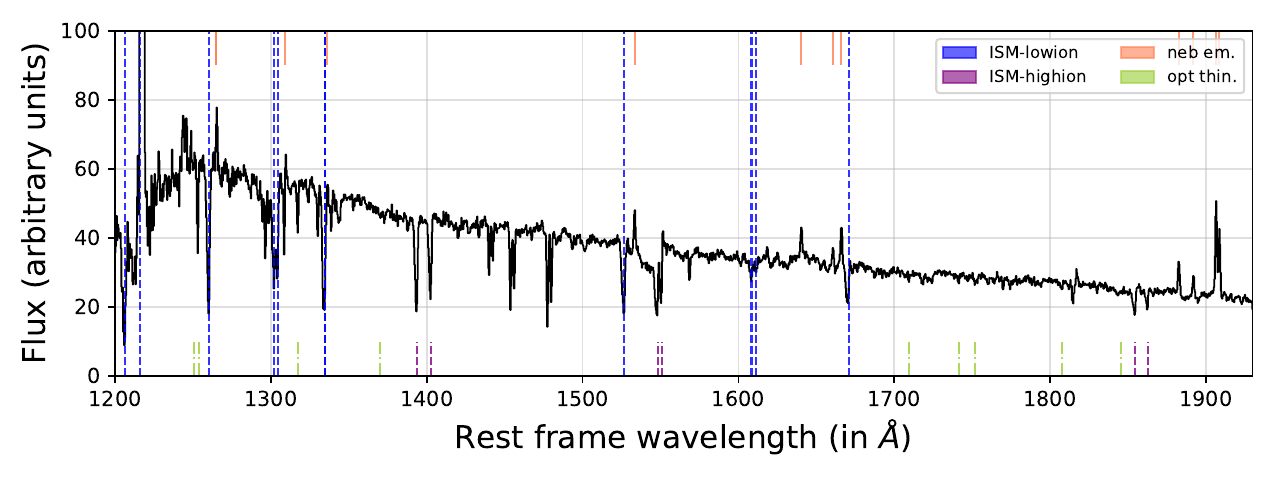} \\
    \includegraphics[width=0.35\linewidth]{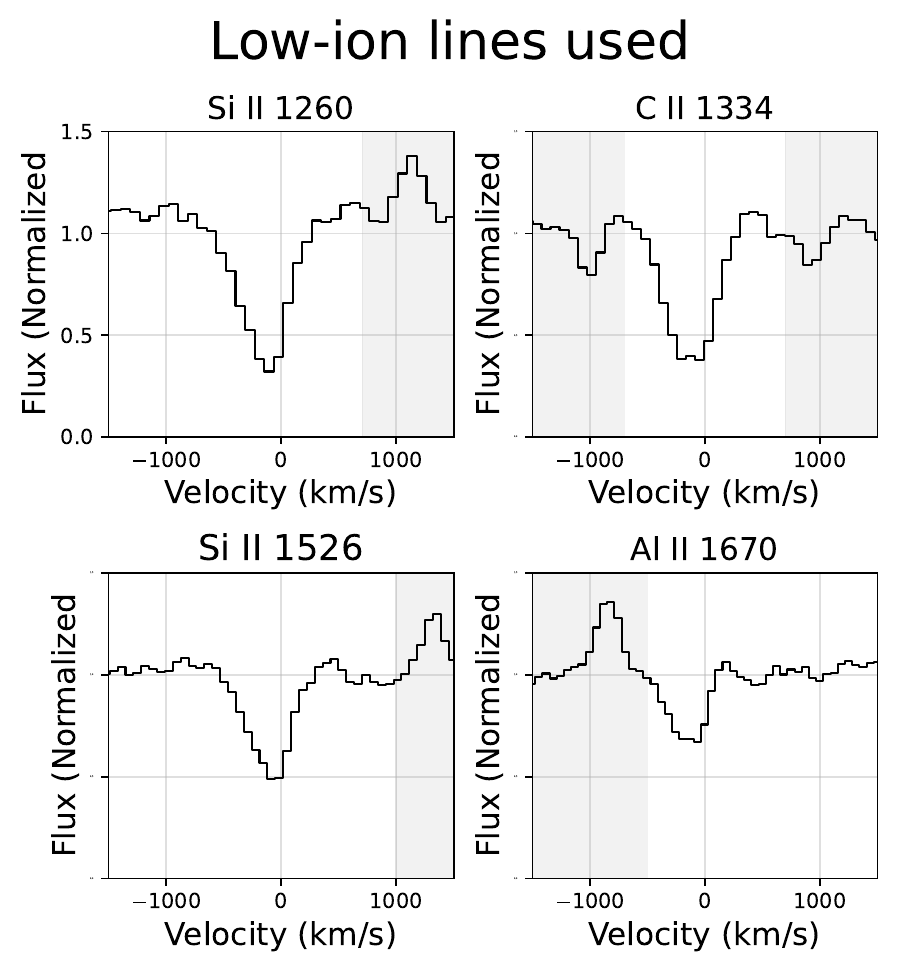} 
    \raisebox{0.5cm}{\includegraphics[width=0.63\linewidth]{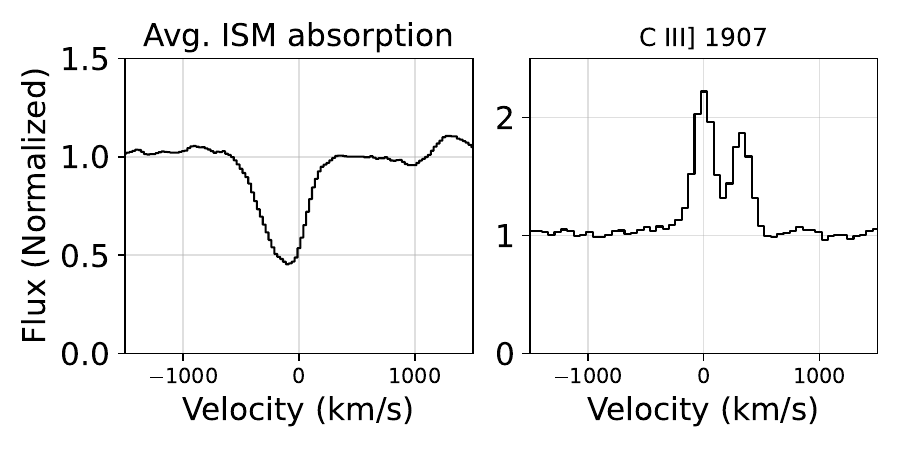}}\\
    \hspace{0.1\linewidth}\Large{(b)} \hspace{0.3\linewidth} \Large{(c)} \hspace{0.24\linewidth} \Large{(d)}\\
    \caption{{(a):} Flux-calibrated KCWI spectrum of CSWA13 obtained by summing the flux from all spaxels covered by the arc. Prominent low ionization (e.g., \ion{Si}{2}~$\lambda$1260, $\lambda$1526, \ion{Fe}{2}~$\lambda$1608, \ion{Al}{2}~$\lambda$1670), high ionization (e.g., \ion{C}{4}~$\lambda\lambda$1549,1551) and optically thin absorption lines (e.g., \ion{Si}{2}~$\lambda$1808, \ion{Ni}{2}~$\lambda$1317, ~$\lambda$1370, ~$\lambda$1709, ~$\lambda$1741)  are marked in blue, purple, and green respectively. Fine structure emission from \ion{Si}{2}* and nebular emission lines (e.g., \ion{C}{3}]) are marked in coral.  {(b):} Velocity profiles of various low-ion ISM absorption lines (from \ion{Si}{2}, \ion{C}{2} and \ion{Al}{2}) used in this study. The gray shaded regions are masked and not used for ISM absorption analysis due to the presence of other significant features. For example, \ion{C}{2}~$\lambda$1334 is flanked by absorption lines from intervening systems which are masked out. Similarly, for \ion{Al}{2} the region affected by nebular \ion{O}{3}]~$\lambda$1666 emission is also excluded from the absorption line analysis. Panel (c) shows the average ISM absorption profile obtained from combining the different low-ionization absorption profiles shown in panel (b). We use this combined absorption profile to probe the outflow kinematics across the arc (described in Section~\ref{subsec:measure-v50}). The lower-right panel (d) shows strong \ion{C}{3}] nebular emission which we use to trace the systemic velocity field across the arc.
    }
    \label{fig:spectra-arc}
\end{figure*}

Galaxies self-regulate their growth across cosmic time through processes of gas outflows, inflows, and recycling \citep[e.g.,][]{Dave2011,Lilly_bathtubmodel}. As stars form and evolve in a galaxy, they inject energy and momentum into the surrounding gas through feedback processes such as stellar winds and supernovae, which in turn redistribute and enrich the interstellar and circumgalactic medium (ISM and CGM) with metals \citep[e.g.,][]{peroux2020}. Surveys of high redshift $z\gtrsim2$ star-forming galaxies \citep[e.g.,][]{shapley2003, steidel2010} have detected near-ubiquitous outflow signatures in absorption, and the prevalence of a metal-enriched CGM has been established to $z\sim6$ and beyond using background quasars \citep[e.g.,][]{Becker_quasars_2009}. The primary mechanism attributed to the outward transport of gas and metals is galactic-scale outflows, i.e., gas being expelled across the entire galaxy. 

Theoretical work has suggested that galactic outflows in intermediate mass galaxies at $z\sim2$ have typical mass loading factors $\eta \sim$~1--10 \citep[where $\eta = \frac{\dot{M}_{out}}{\mathrm{SFR}}$, $\dot{M}_{out}$ is the mass outflow rate, and SFR is the star formation rate; e.g.,][]{muratov2015,muratov2017,tng50,pandya2021}. Chemical evolution analysis supports similarly high mass loading \citep[e.g.,][]{sanders2021}.
However, directly measuring the gas mass loss rates (and mass loading factors) from galaxies has been challenging due to uncertainties such as ionization corrections, metallicity, and especially the radial extent of outflows which requires spatially resolved measurements.
Characterizing galaxies at high redshifts requires sophisticated instruments and long integration times on 8--10m class telescopes. Even with such facilities, the ability to conduct spatially resolved observations is limited to the brightest galaxies at high redshifts. Gravitational lensing, whereby distant background galaxies are magnified by massive galaxies along the line of sight \citep[e.g.,][and references therein]{Schneider1992,narayan1996,tomasso-ara-review}, offers a promising way to carry out resolved studies at high redshifts \citep[e.g.,][]{jones2013, Bordoloi2014,nicha-osiris-survey,  spilker2020_moloutflows,Shaban_2022}. 
Dedicated searches for strong lens systems have resulted in substantial and growing samples \citep[e.g.,][]{danstark-cswa-confirmation, VyAGEL}.

Tens of lens systems have now been followed up with deep slit spectroscopy to characterize their outflow properties (e.g., outflow velocity), taking advantage of the lensing magnification \citep{tucker-dustinthewind,rigby-megasura-paper1,kvgc_ESI2022}. 
Recently, integral field spectroscopic (IFS) observations have shown great promise in spatially resolving the outflows in high-redshift galaxies, including direct measurements of their spatial extent \citep{finley2017, burchett2021, Shaban_2022, shaban_massloss}. 
In the case study presented in this paper, we seek to more robustly establish the mass loss rate ($\dot{M}_{out}$) and mass loading factor ($\eta$) of the low-ionization gas phase by combining IFS measurements of \ion{H}{1} column density, outflow velocity, radial extent and the spatial structure of outflowing gas.

Our target is CSWA13, a bright star-forming galaxy discovered as part of the Cambridge And Sloan Survey Of
Wide ARcs in the skY \citep[CASSOWARY;][]{belokurov2009}. The lensing nature of this system was spectroscopically confirmed by \citet{danstark-cswa-confirmation} with redshifts of $z_d=0.41$ for the foreground deflector galaxy and $z_s=1.87$ for the background bright arc.
An analysis of the stellar populations and UV nebular emission of this target is presented by \citet{ramesh_CSWA}. Here, we use integral field spectroscopy from the Keck Cosmic Web Imager \citep[KCWI;][]{KCWI_Morrissey_2018} at the W.~M. Keck Observatory to spatially map the outflowing gas from this galaxy.

This paper is organized as follows. Section~\ref{sec:kcwi-observations} describes the KCWI observations and data reduction. Section~\ref{sec:lensmodel} describes the lens model used to obtain accurate intrinsic properties of the source galaxy. The spectroscopic analysis methodology is described in Section~\ref{sec:methodology}. We discuss the results in Section~\ref{sec:results-and-discussion} with conclusions in Section~\ref{sec:conclusions}. Throughout this work, we assume a flat $\Lambda$CDM cosmology with $H_0 = 70$ km s$^{-1}$ Mpc$^{-1}$ and $\Omega_{\rm m} = 0.3$.

\section{KCWI observations}\label{sec:kcwi-observations}

We observed CSWA13 with KCWI on 06 April 2022 with the BL grating and Medium slicer configuration. The total wavelength coverage is 3229--5825~\AA\ corresponding to 1127--2032~\AA\ in the $z=1.87$ galaxy rest frame, with spectral resolution of 2.4~\AA\ FWHM ($R\simeq1800$). The total exposure time was 2 hours, with six exposures of 1200 seconds each. Three exposures were taken at each of two orthogonal position angles (PA) of 0 and 90 degrees, both of which covered the entire lens system (Figure~\ref{fig:CSWA13-system-intro}). The observing conditions ranged from clear to thin clouds with seeing of $\sim1-1\farcs3$ FWHM. For sky subtraction, a nearby blank sky reference area was observed for 300 seconds with each set of PA exposures.

The data were reduced using the IDL version of the KCWI data reduction pipeline (KDERP-v1.0.2)\footnote{\url{www.github.com/Keck-DataReductionPipelines/KcwiDRP}}. We initially ran stages 1--4 of the pipeline, which perform bias, gain, and dark current corrections as well as cosmic ray rejection. We use the observed sky frame for sky subtraction in stage 5 of the pipeline, as there is insufficient blank sky background in the lensing field of CSWA13 itself. Flux calibrated and DAR (differential atmospheric refraction) corrected datacubes are obtained by running stages 6--8.
Flux calibration was carried out using observations of the standard star BD26D2606 taken on 20 June 2020 using the same instrumental setup. The flux calibration is uncertain due to the non-photometric conditions, but the analysis and results of this paper do not require absolute flux calibration. We correct for any nonzero residual background in each wavelength slice in the following way: we consider the median flux within a 48~\AA\ wavelength bin centered on each slice, mask out regions containing detected sources, fit the resultant background profile with a 2D first-order polynomial, and remove its contribution from the 2D slice. This is similar to the procedure described in \citet{Mortensen2021}, which was applied to a narrow wavelength range, while here we are interested in features that span the entire wavelength range of the KCWI spectra. The processed datacubes were then resampled to a common grid, aligned, averaged, and rebinned to a spatial pixel size of 0\farcs3$\times$0\farcs3.
We also converted the pipeline output from air to vacuum wavelengths. We use the $1\sigma$ standard deviation of flux from blank regions of the sky to estimate the uncertainty of our spectra. 

Figure~\ref{fig:CSWA13-system-intro} shows a color composite image generated using HumVI \citep{phil-humvi} from the reduced KCWI data cube by summing three broadband wavelength regions centered at 3730, 4430 and 5130 \AA, each with a width of 400 \AA. Multiple images of the arc are clearly identifiable by their blue color in the KCWI data, as well as prominent spectral features. Figure~\ref{fig:CSWA13-system-intro} \emph{(Left panel)} also shows the Hubble Space Telescope (HST) near-infrared image obtained with WFC3-IR using the F140W filter. We identify three distinct star-forming complexes in the HST image (A, B, and C), which are multiply imaged across the entire arc. The integrated KCWI spectrum from summing spaxels containing the arc is shown in Figure~\ref{fig:spectra-arc}(a) for the rest-frame wavelength range 1200--1930~\AA. Metal absorption lines from three intervening galaxy systems at $z=0.81, 1.67, 1.69$, $1.74$ and $1.76$ are also detected in the arc spectra, with prominent \ion{C}{4} and other features. We mask their contribution in all the analyses presented in the rest of this paper.

\section{Lens Model}\label{sec:lensmodel}
Gravitational lens modeling is necessary to reconstruct the galaxy's source plane morphology and intrinsic properties. 
We use the lens modeling software \textsc{lenstronomy} \citep{Birrer18, Birrer21b} to build the lens model. We adopt the conjugate point modeling method which is commonly used in cluster-scale lens modeling \citep[see][and references therein]{Kneib11}, with multiple image positions for the A, B, and C components of the source (Figure~\ref{fig:CSWA13-system-intro}). 
Our KCWI data confirm the multiple image nature of components A, B, and C, for example via their distinct velocity structure and \Lya\ profiles.
The lens model is optimized with the nested sampler \textsc{dynesty} \citep{Speagle19}. Our lens model consists of an elliptical Navarro--Frenk--White \citep[NFW;][]{Navarro96, Navarro97} halo profile for the lensing galaxy group, a double Chameleon profile for the stellar mass distribution of the central galaxy \citep{Dutton11}, a point mass to account for the supermassive black hole (SMBH) at the center of this galaxy, eight singular isothermal ellipsoid profiles to model the eight brightest galaxies within 20 arcsec of the central galaxy \citep{Kassiola93}, a residual shear field and a residual flexion field.

The ellipticity of the projected NFW halo is parameterized in the convergence or the surface density and not in the potential \citep{Oguri21}. The Chameleon profile is a combination of two non-singular isothermal ellipsoids that provide a good approximation to the S\'ersic profile within $0.5 - 3R_{\rm eff}$, where $R_{\rm eff}$ is the half-light radius \citep{Sersic68, Dutton11}. We find that a superposition of two S\'ersic profiles is necessary to fit the light distribution of the central galaxy well \citep{Claeskens06, Suyu13}. We convert the best-fit parameters of the double S\'ersic profile into the parameters of the double Chameleon profile. The pre-fitted parameters determining the angular and radial shape of the double Chameleon profile are fixed during the lens model optimization, and only the mass-to-light ratio is free. We fix the NFW halo mass to be $10^{14} M_{\odot}$ and impose a Gaussian prior on the concentration parameter ($c_{200} = 5.0 \pm 0.8$) following the results of \citet{Newman15} for group-scale halos. The ellipticity and centroid of the NFW halo are fixed to be the same as the ellipticity and centroid, respectively, of the central galaxy's light distribution. Thus, the concentration is the only free parameter for the NFW profile. We impose a prior on the stellar mass to SMBH mass relation using the results from \citet{Li23}.

\begin{figure}
    \centering
    \includegraphics[width=0.95\columnwidth]{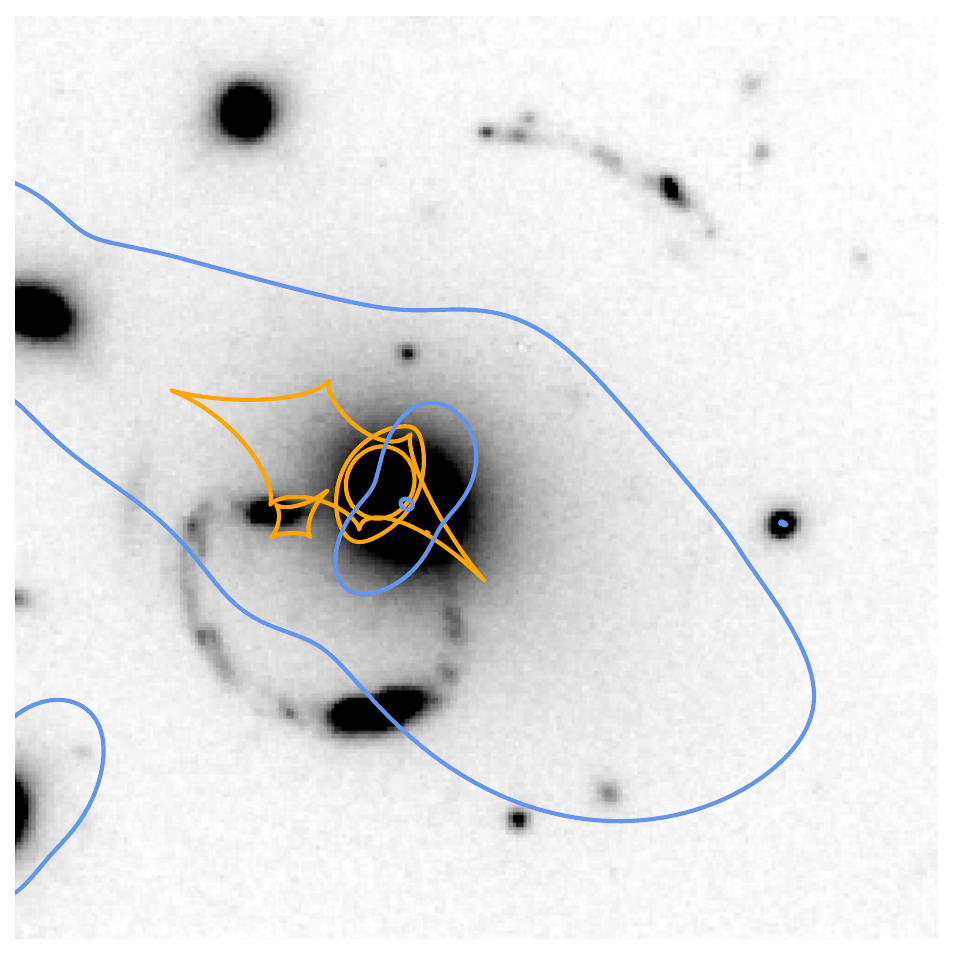}\\
    \includegraphics[width=0.49\columnwidth]{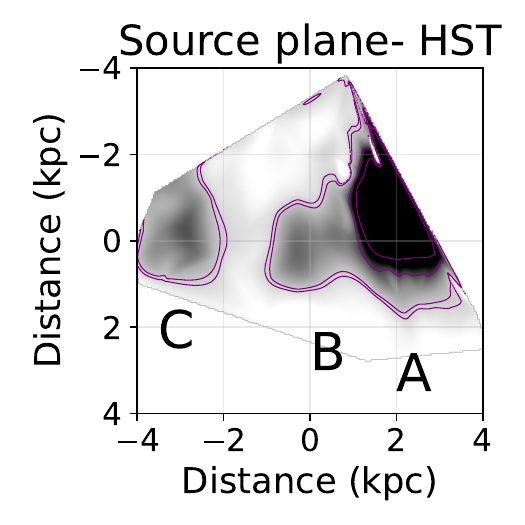}
    \includegraphics[width=0.49\columnwidth]{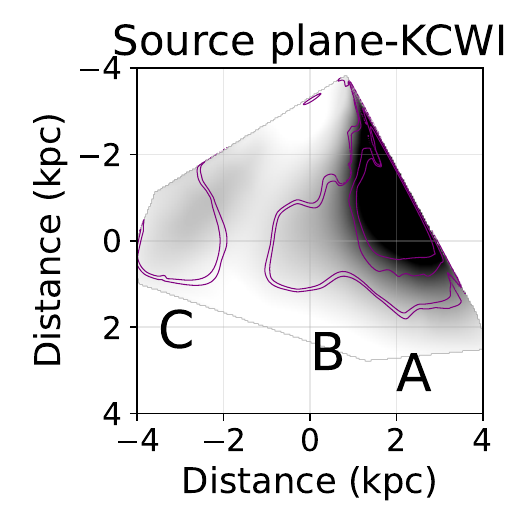}    
    \caption{\emph{Top}: Illustration of the best-fit lens model (described in Section~\ref{sec:lensmodel}). The critical and caustic curves (which denote regions of maximal magnification in the image and source planes) are shown in blue and orange respectively.
    \emph{Bottom}: Source plane reconstructed maps of the HST and KCWI continuum images, revealing the clumpy galaxy morphology of CSWA13. 
    Purple lines denote the contours obtained from the source plane HST imaging. The three distinct star-forming regions A, B and C span a physical distance of $\sim8$ kpc in the source plane.
    } 
    \label{fig:lens_model1}
\end{figure}

\begin{figure*}
    \centering
    \begin{tabular}{ccc}
        \Large{\ion{C}{3}]}  & \Large{Image Plane} & \Large{Source plane}\\
        \includegraphics[width=0.28\linewidth]{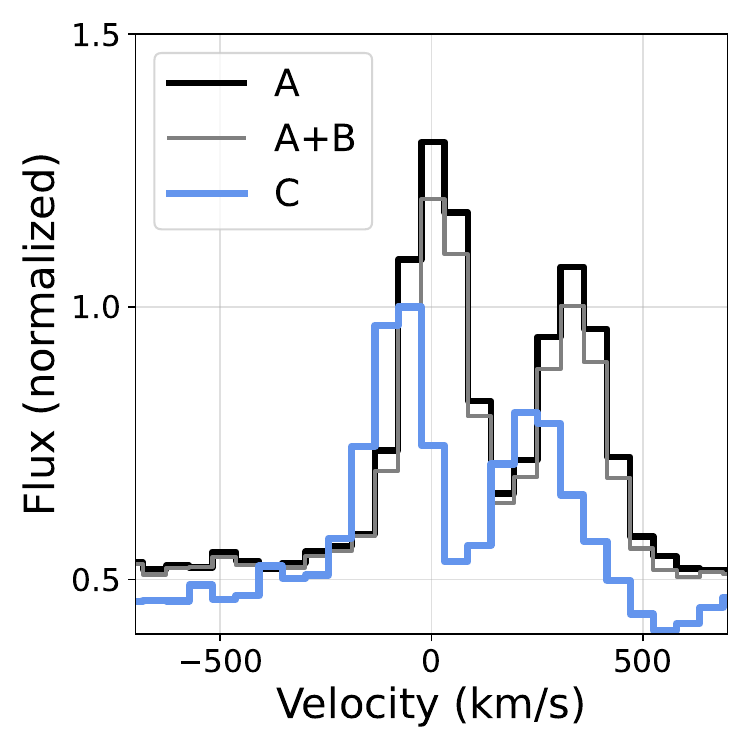}  & \raisebox{0.4cm}{\includegraphics[width=0.6\columnwidth]{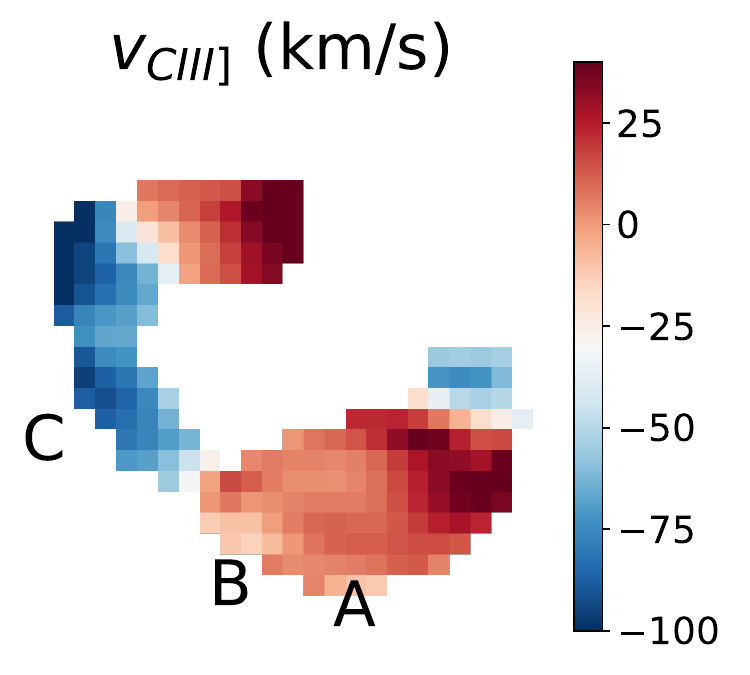}} & \includegraphics[width=0.725\columnwidth ]{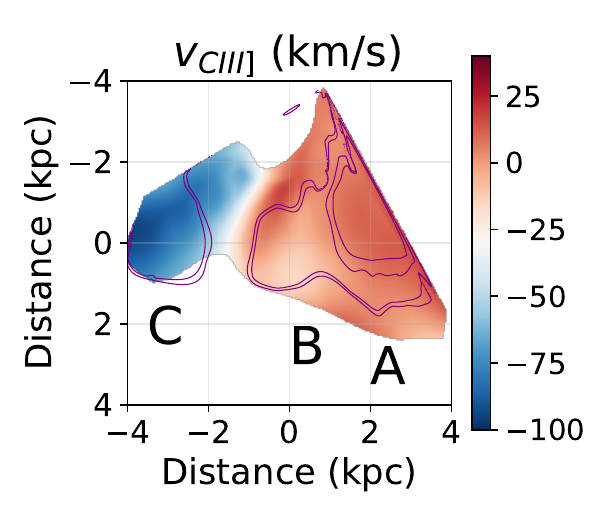} \\        
       \Large{\ion{Si}{2}} &  & \\
      \includegraphics[width=0.3\linewidth]{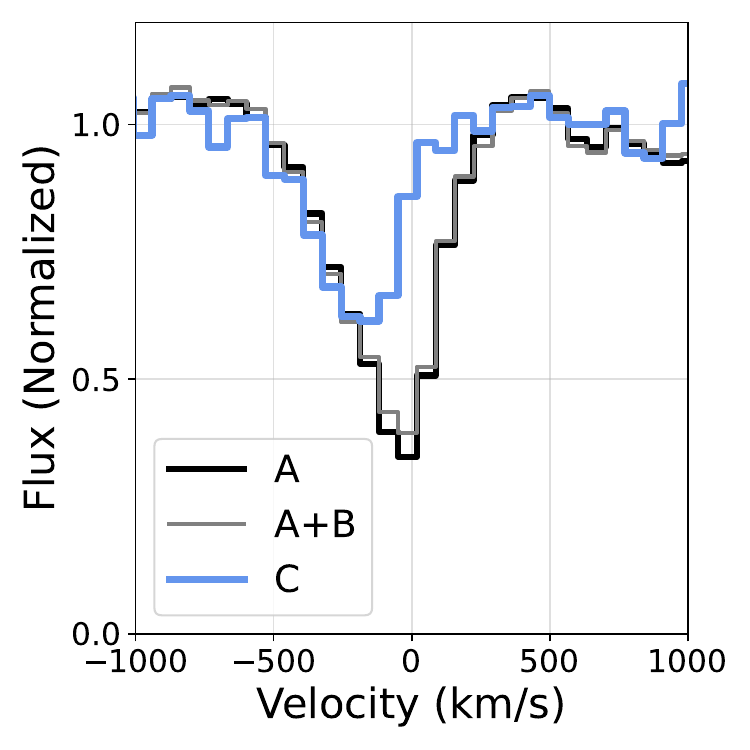}  & \raisebox{0.4cm}{\includegraphics[width=0.6\columnwidth]{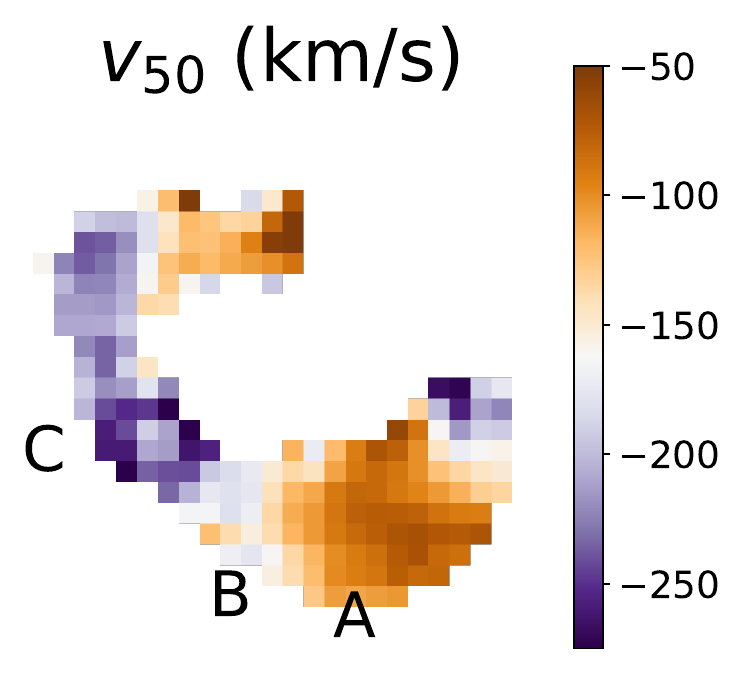}} & \includegraphics[width=0.725\columnwidth ]{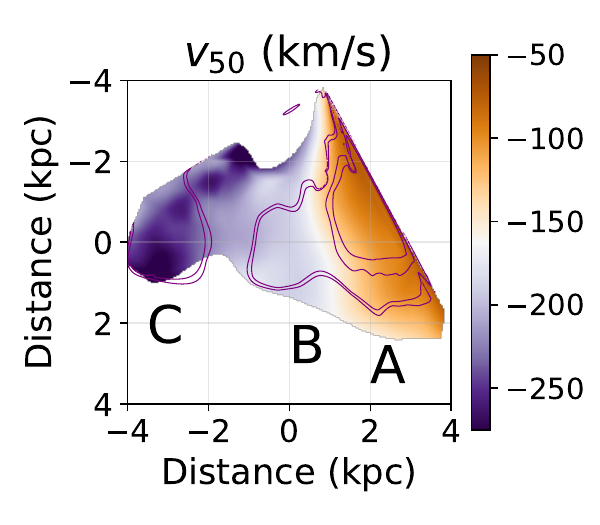} \\         
    \end{tabular}

    \caption{ 
    \emph{Left}: Zoom-in of the \ion{C}{3}]$\lambda\lambda1907,09$ emission and \ion{Si}{2}~$\lambda 1526$ ISM absorption line profile in regions A, A+B and C. \emph{Middle}: Spatial maps of the centroid velocity of nebular \ion{C}{3}] emission which traces the young stars, and the outflow velocity $v_{50}$. The outflow velocity maps are measured with respect to a systemic redshift of  $z_{sys}=\srcRedshiftNeb$ obtained from a galaxy-integrated spectrum. A similar velocity structure is apparent in both, suggesting that the outflowing gas is closely associated with the stars within each $\sim$ kpc spatial resolution element. The fold-symmetry of the velocity fields in the image plane is a result of the multiple image lensing configuration (Figures~\ref{fig:CSWA13-system-intro}, \ref{fig:lens_model1}). \emph{Right}: Source plane maps of the centroid velocity of nebular \ion{C}{3}] emission and outflow velocity $v_{50}$. The purple lines show contours from HST imaging, as in Figure~\ref{fig:lens_model1}. The ISM absorption profiles show significant spatial variation, and we can clearly see that the outflow velocity mirrors the systemic velocity structure along the major axis. We discuss the quantitative comparison of outflow and nebular velocity and its implications in Section~\ref{subsec:young-outflows}.   
    } 
    \label{fig:sourceplane-velmaps}
\end{figure*}

For the eight nearby galaxies included explicitly in the model, we use the \textsc{photutils} package to measure aperture photometry and ellipticity. We fix the centroid and ellipticity of these galaxies and only allow the Einstein radii of each galaxy as free parameters. The residual shear (also called ``external shear'' in the literature) has two free parameters (i.e., the shear magnitude $\gamma_{\rm shear}$ and angle $\phi_{\rm shear}$) and the residual flexion field has four parameters. These residual shear and flexion fields account for both the ``internal'' angular structure of the central lensing galaxy that is not fully captured by the ellipticity parametrization of the NFW and Chameleon profiles, and also ``external'' contributions from mass structure beyond the galaxies, which are explicitly accounted for in our model. The residual flexion field is also necessary to correctly reproduce the atypical image configuration of the multiple components in this system. This residual flexion is plausible since many group-member galaxies are not accounted for in our lens model except for the brightest eight. Our model has 17 free parameters in total, and there are 24 data points from 4 positions for each of the A, B, and C image sets. The lens model is optimized by minimizing the total distances between the mapped positions of each image set on the source plane. 

Figure~\ref{fig:lens_model1} illustrates the best-fit lens model with the caustic and critical curves. The lower panels of Figure~\ref{fig:lens_model1} show the intrinsic galaxy morphology from HST and KCWI continuum images, reconstructed in the $z=1.87$ source plane. The three star-forming complexes (A, B and C) span a physical distance of $\sim8$ kpc in the source plane, using our fiducial cosmology. We find that the mean areal magnification of the southwestern counter-image of the galaxy is $|\mu|=\MeanMagValue$. We obtain a magnification corrected stellar mass and SFR of $\log_{10}(M_*/\Msun)=\StelMassValue$ and $\log_{10}({\rm SFR/(\Msun~yr^{-1}}))=\SFRValue$, after scaling the BEAGLE outputs presented in \citet{ramesh_CSWA}.

\begin{figure*}[ht!]
    \centering
    \includegraphics[width=\linewidth]{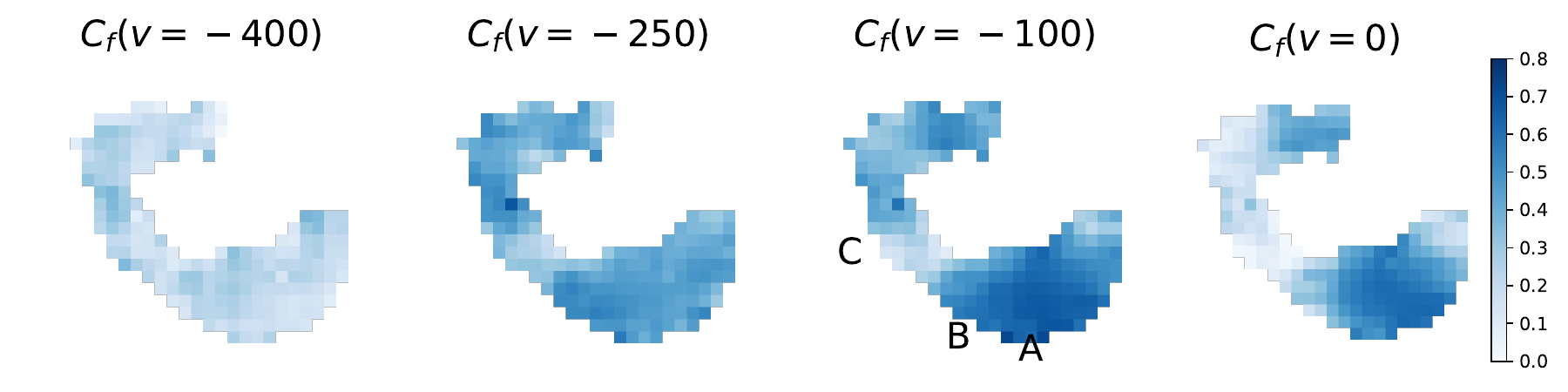}  \\    
    \includegraphics[width=\linewidth]{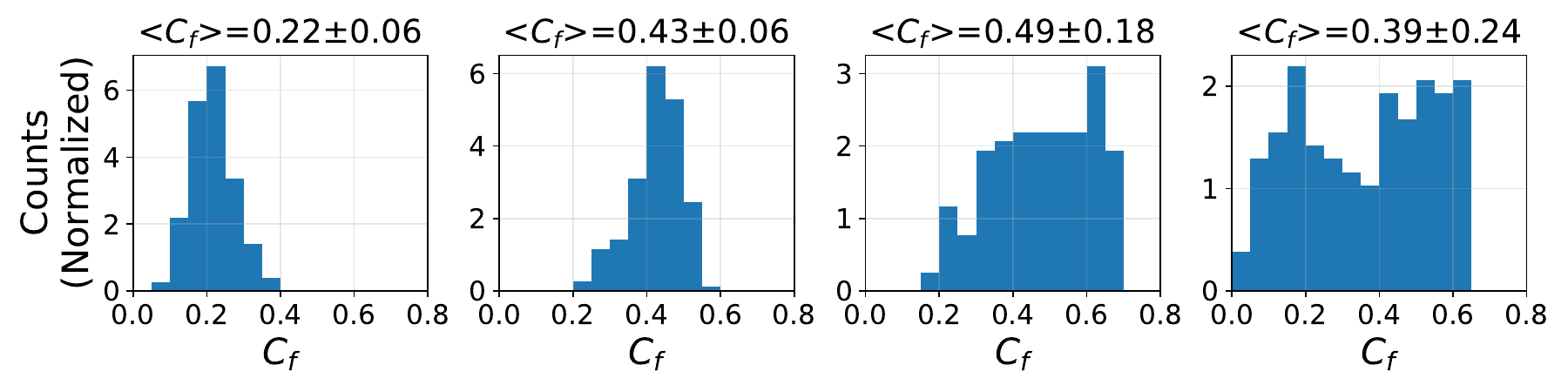}\\

    \caption{\emph{Top}: Spatial map of the covering fraction $C_f$ of the low ionization gas at outflow velocities of $v=-400$ to $-100$~\kms\ along with the systemic velocity ($v=0$~\kms). \emph{Bottom}: Distribution of $C_f$ in each velocity bin. Gas at higher outflow velocities is more uniformly distributed across the entire galaxy, for example with a mean $C_f(v=-250)=0.43$ with a relatively small scatter of 0.06. However, at systemic velocities it is more heterogenous, with a substantial variation in the covering fraction of gas with region A and B having $C_f\sim0.6$ compared to $C_f\sim0.1$ in region C. We also find that regions A+B have blueshifted \Lya\ absorption and redshifted emission (Figure~\ref{fig:lya-cswa13}) whereas region C has broad \Lya\ emission extending to blueshifted velocities. Therefore, regions A+B likely have a higher column density of slow-moving gas at systemic velocities seen in metal absorption transitions as well as \Lya, with a paucity of gas at $v\sim0$ in region C.
    }
    \label{fig:coveringfraction-cswa13}
\end{figure*}

\subsection{Galaxy morphology}
CSWA13 is a moderately dusty (with V-band dust attenuation $A_V \simeq 0.6$; \citealt{ramesh_CSWA})
galaxy. Its specific SFR $=43.1^{+56.6}_{-28.7}$ Gyr$^{-1}$ places it above the star-forming main sequence at $z\sim2$ \citep[e.g.,][]{z2_mainsequence}. The spatially resolved kinematics from this work uniquely allows us to tie the observed morphology with the kinematics of the gas around the galaxy. The clumpy morphology of CSWA13 (Figure~\ref{fig:lens_model1}) resembles a tadpole or a chain galaxy \citep[e.g.,][]{cowie-1995-HST,sidney-1996-hubbledeepfield,Elmegreen_2005,forster-schreiber-SINS-survey,elmegreen-tadpolegalaxies-udf} with a bright head (region A) and a tail (regions B and C). These tadpoles are common at $z\sim2$, comprising $\sim$10\% of the galaxies in the Hubble Ultra Deep Field (HUDF) and $\sim$44\% if these tadpoles are indeed edge-on projections of the clump-cluster \citep{Elmegreen_chains_and_disks} and chain galaxies.  While the nebular kinematics for a handful of tadpole galaxies at $z\sim2$ were measured as part of the SINS survey \citep{forster-schreiber-SINS-survey} and show similar nebular kinematics as CSWA13, this paper is the first study to simultaneously map nebular and outflow kinematics in a galaxy of this type.

\begin{figure*}
    \centering
     \vspace{-0.1cm}
    \includegraphics[width=0.44\linewidth]{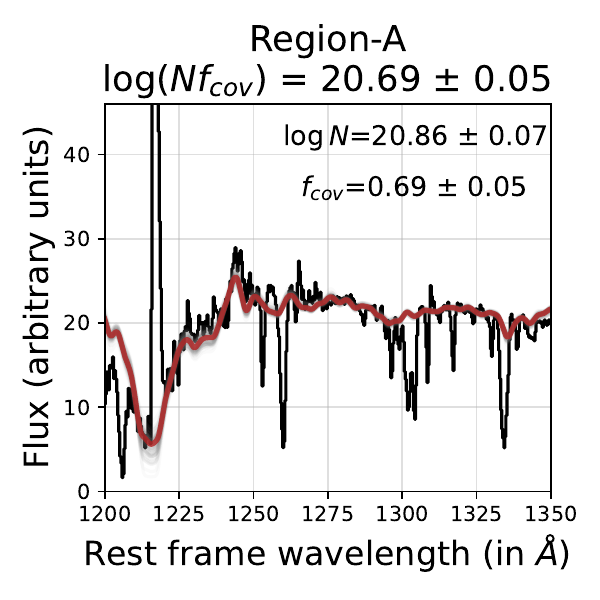}      
    \includegraphics[width=0.44\linewidth]{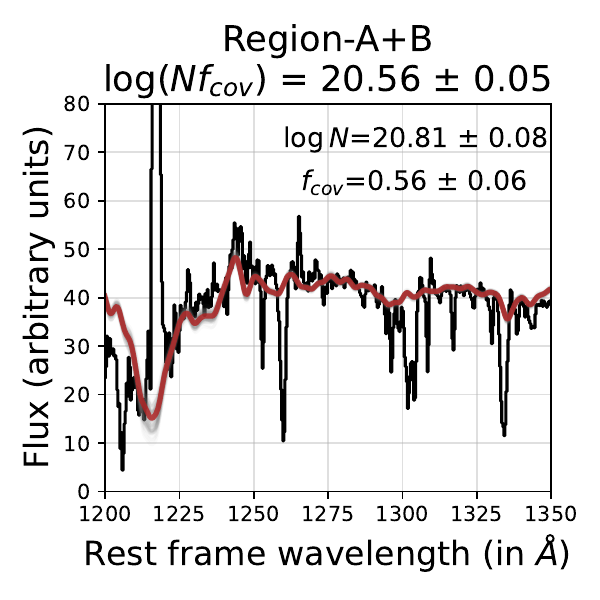}  \\   
    \includegraphics[width=0.44\linewidth]{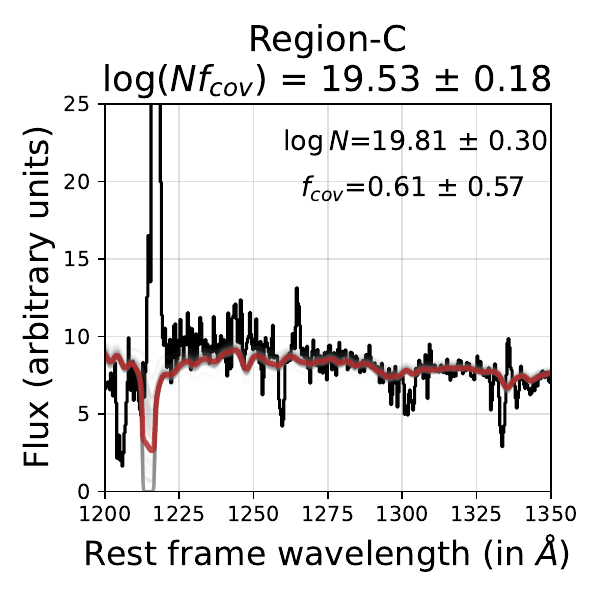}   
     \includegraphics[width=0.42\linewidth]{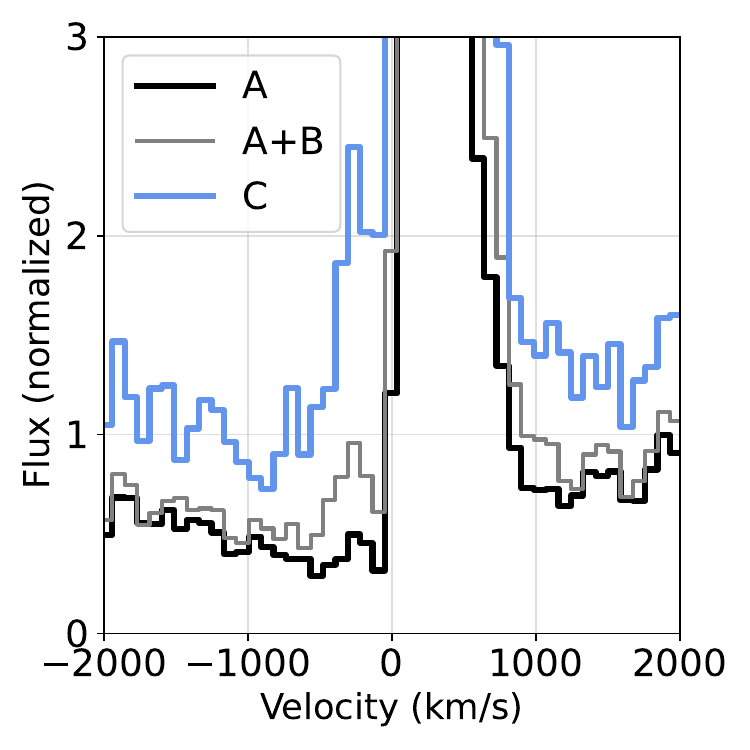}
     
    \caption{Spectra of regions A, A+B and C (black) along with the best-fit Voigt and stellar population fit (described in Section~\ref{subsec:voigt-fit}) shown in red. The gray lines show the 100 independent realizations used to obtain the best fit. \emph{Bottom Right}: Close-up of the \Lya\ profile showing differences in blueshifted emission ($v < 0$~\kms) from each region. Region A+B has higher column density and covering fraction ($\log N\sim 21$; $f_{cov}\sim0.6$) compared to Region C. 
    This is complemented by our spatial maps of covering fraction ($C_f(v)$) obtained by independently analyzing the ISM metal absorption lines (Figure~\ref{fig:coveringfraction-cswa13}) which show a similar variation in $C_f$ across the galaxy.
    }
    \label{fig:lya-cswa13}
\end{figure*}

\begin{figure*}
    \centering
    \hspace{-1.5cm}
    \includegraphics[width=0.355\linewidth]{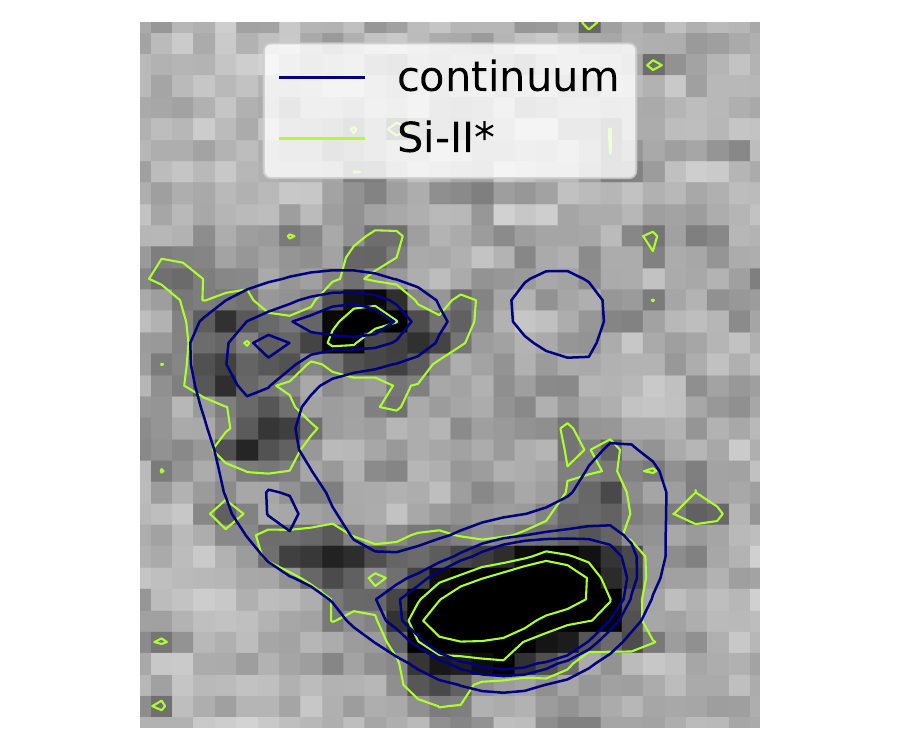}\hspace{-1cm}
    \includegraphics[width=0.3\linewidth]{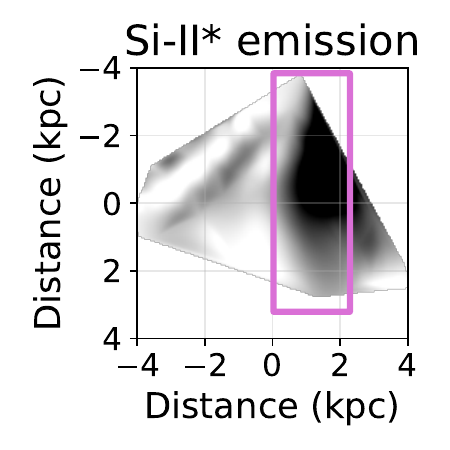}
    \includegraphics[width=0.3\linewidth]{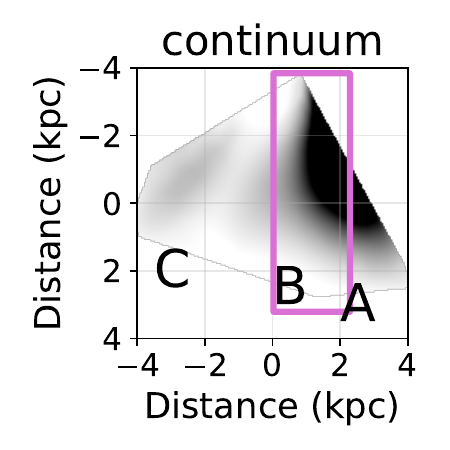}\\
    \vspace{1cm}
    \includegraphics[width=0.42\linewidth]{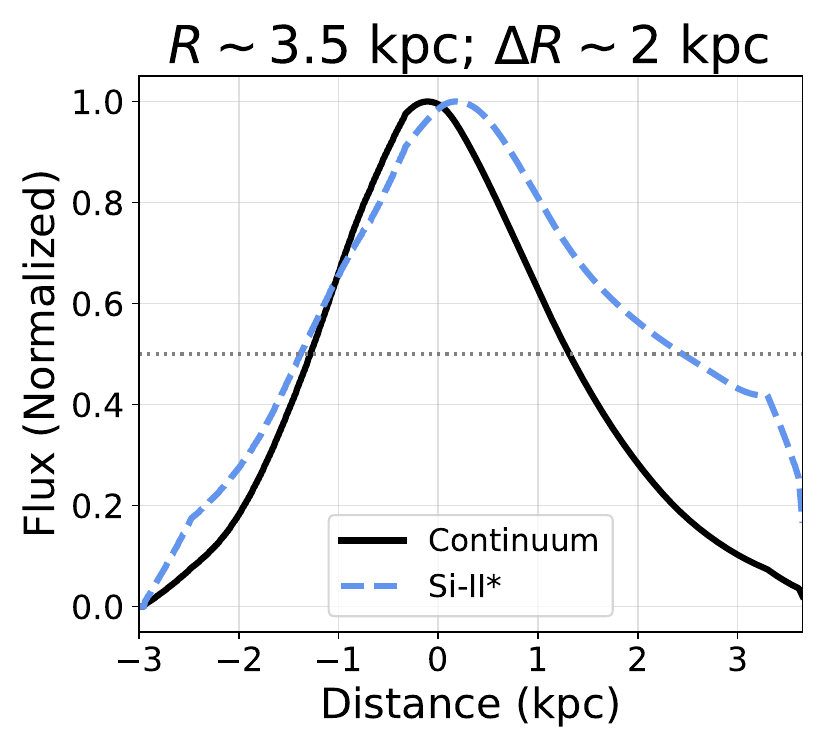}
    \raisebox{0.cm}{\includegraphics[width=0.38\linewidth]{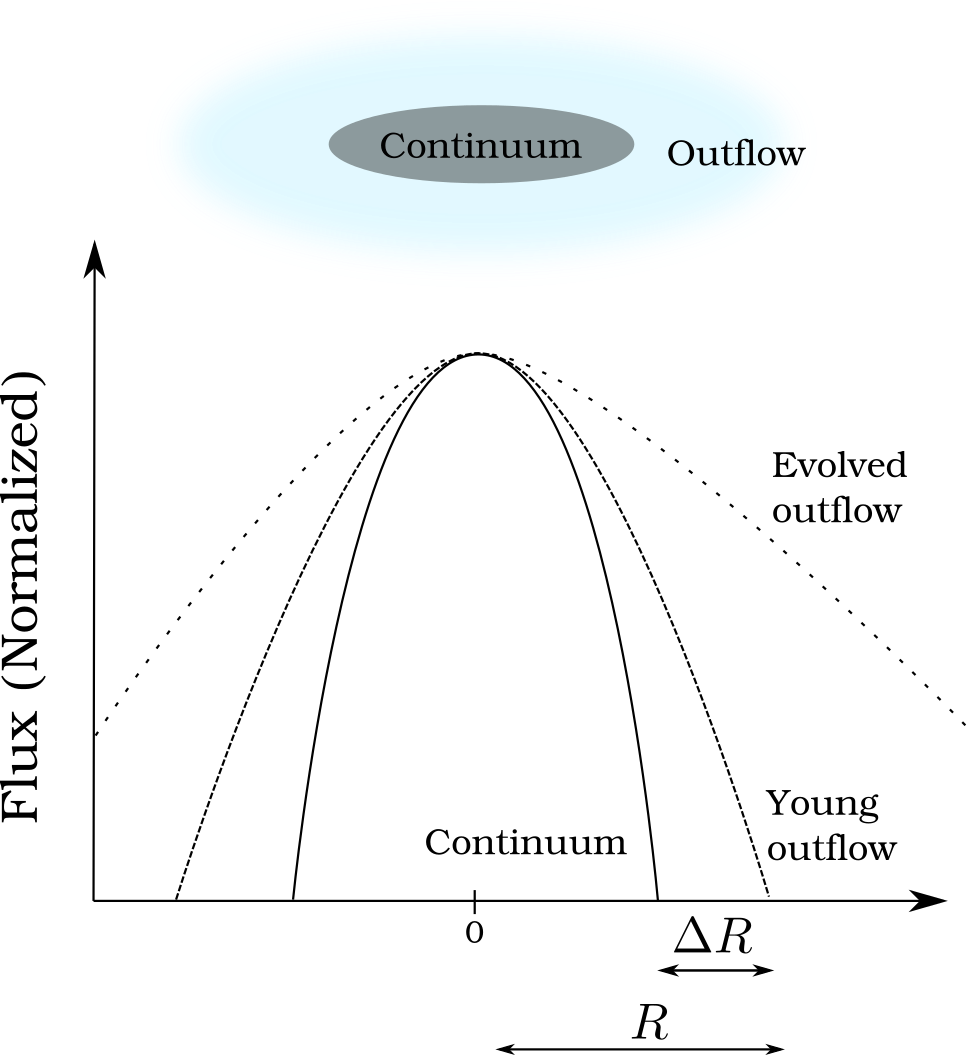}}    
    \caption{
    \emph{Top Left}: Image plane \ion{Si}{2}*~$\lambda$1533  emission line map obtained from a single Gaussian fit (described in Section~\ref{subsec:emission-linemap}). Contours of the continuum and \ion{Si}{2}* emission are shown in blue and green respectively. We find that the fluorescent emission is spatially extended compared to the continuum in all the three fold images of the arc. \emph{Top Right}: Source plane continuum and \ion{Si}{2}*~$\lambda$1533 emission line maps. The purple box shows the position of a pseudo slit used to extract the continuum and emission line flux. It is centered on regions A+B which show the highest column density in absorption, and aligned roughly along the kinematic minor axis (see Figure~\ref{fig:sourceplane-velmaps}).  
    \emph{Bottom Left}: Plot of extracted continuum and \ion{Si}{2}* flux from the pseudo slit as a function of distance, showing larger spatial extent of fluorescent \ion{Si}{2}* compared to the stellar continuum. \emph{Bottom Right}: Illustration of two different outflow scenarios (young outflow, evolved outflow) in this phase space. $R$ measures the radial extent of the outflowing gas from the center of the galaxy. $\Delta R$ in the spherical shell model represents the thickness of the shell that the outflow is enclosed in. Here we assume that the center of the galaxy is at the center of pseudo slit and define $\Delta R = R - R_{cont,50}$ where $R_{cont,50}$ is the radii where the flux reaches 50\%. Based on this nomenclature, we estimate $R\sim3.5$~kpc and  $R/\Delta R\sim2$.
    }\label{fig:fineemission}
\end{figure*}

\section{Spatially resolved ISM gas kinematics}\label{sec:methodology}

Our overall aim is to quantify the key properties of outflows in CSWA13 such as the spatial extent, mass loss rate, and mass loading factor. This in turn requires measuring various physical quantities which we address in this section: spatial maps of outflow velocity ($v_{out}$), covering fraction and solid angle ($\Omega$), \ion{H}{1} column density ($N$), and the radial extent of outflowing gas ($R$). 
Ultimately the spatially resolved KCWI spectroscopy provides a comprehensive view of the warm ($T\sim10^4$~K) outflowing gas, including the mass loss rate (e.g., $\dot{M} \propto v_{out}\Omega N R$), as we will discuss in Section~\ref{sec:results-and-discussion}.

\subsection{Systemic redshift and velocity field}\label{subsec:neb-kinematics}

For spatial analysis of the outflow kinematics, we require not only the systemic redshift but rather the velocity field of young stars in the galaxy. 
We derive this using the nebular \ion{C}{3}]$\lambda\lambda1907,09$ emission line doublet, which traces the dense ISM gas in \ion{H}{2} regions associated with newly-formed stars. \ion{C}{3}] is the most pragmatic tracer of the velocity field as it has a high signal-to-noise ratio relative to stellar absorption features.
We fit a double Gaussian function to the \ion{C}{3}] doublet assuming the same redshift and velocity dispersion for both lines, and  obtain a systemic redshift $z_{sys}=\srcRedshiftNebErr$ for the integrated galaxy spectrum. 

The best-fit velocity centroid in each spaxel relative to this systemic redshift is shown in Figure~\ref{fig:sourceplane-velmaps}, revealing a coherent velocity shear of $\sim$150~\kms\ seen consistently in all four multiple images.

A fit to the integrated stellar photospheric absorption lines \ion{Si}{3}~$\lambda$1294,1298 yields a systemic redshift of $z_{sys}=\srcRedshiftErr$, which agrees with \ion{C}{3}] within the measurement uncertainties.
We note that the difference between stellar and nebular systemic redshift corresponds to a $\sim 2\%$ change in the measured outflow velocity ($v_{50}$), which does not affect our results. 
A detailed analysis of the stellar kinematics will be presented in a future work (Rhoades et al. in prep). We use the redshift obtained from the integrated spectrum for measuring the outflow velocities ($v_{50}$) in Section~\ref{subsec:measure-v50} and discuss the spatially resolved outflow kinematics (i.e., $v_{50} - v_{sys}$) further in Section~\ref{subsec:young-outflows}.

\subsection{Outflow velocity and covering fraction}\label{subsec:measure-v50}

We follow the methodology described in \citet{kvgc_ESI2022} to analyze the ISM absorption profiles and derive outflow velocity metrics. For each spaxel, we obtain a covering fraction profile ($C_f$) by normalizing the spectra and combining the flux from \ion{Si}{2}~$\lambda 1260$,  \ion{C}{2}~$\lambda 1334$, \ion{Si}{2}~$\lambda 1526$ and \ion{Al}{2}~$\lambda 1670$ using an inverse variance weighted average. 
We mask regions in the spectra which are affected by intervening absorption systems (e.g., $-$1000 and 1000 \kms\ from \ion{C}{2}) and nebular emission features (e.g., \ion{O}{3}] blueward of \ion{Al}{2}). 
Figure~\ref{fig:spectra-arc}'s bottom panels demonstrate our methodology applied to the integrated galaxy spectrum. We note that the low-ion ISM absorption lines seen in Figure~\ref{fig:spectra-arc} as well as in individual spaxels show a clear asymmetric profile, with a blueshifted velocity centroid and an absorption wing extending to outflow velocities $v\gtrsim250$ \kms. Thus, we use a double Gaussian function to fit the resulting mean absorption line profile, which adequately captures the skewness apparent in the absorption line \citep{kvgc_ESI2022}. The covering fraction in each spaxel can be parameterized in the following form for ISM absorption lines with high optical depths ($\tau \gg 1$):
\begin{equation}
    \frac{I}{I_0}(v) = 1 - C_f(v)
\end{equation}
\begin{equation}
    C_f(v) = C_{f,G1}(v) + C_{f,G2}(v)
\end{equation}
where $I_{0}$ is the stellar continuum, $I/I_0$ is the normalized line flux as is shown in the lower left panel of Figure 4 , and $C_{f,G1}$ and $C_{f,G2}$ are Gaussian functions which capture the faster- and slower-moving velocity components respectively.

From the fitted profiles, we measure the velocity centroid $v_{50}$ (defined as the 50th percentile of absorption equivalent width) which traces the bulk outflow motion in the galaxy. The $v_{50}$ metric is also robust to resolution and blending effects \citep[as discussed in ][]{kvgc_ESI2022}. Figure~\ref{fig:sourceplane-velmaps} shows the $v_{50}$ map obtained for all spaxels in the arc which have a continuum signal-to-noise ratio (SNR) $>4$ per pixel at representative wavelengths, enabling reliable fits to the ISM absorption. The $v_{50}$ maps show that the bulk outflow velocity of the ISM gas varies across different star-forming complexes in the galaxy, from $|v_{50}|\lesssim 100$~\kms\ in region A to $|v_{50}|\gtrsim 200$~\kms\ in region C, relative to the adopted $z_{sys}=\srcRedshiftNeb$. The spatial structure of $v_{50}$ from outflows is however quite similar to the nebular emission velocity (Section~\ref{subsec:neb-kinematics}). The outflow velocity relative to \textit{local} systemic redshift is thus relatively constant. We discuss the implications of this in detail in Section~\ref{subsec:young-outflows}.

Figure~\ref{fig:sourceplane-velmaps}~(left panel) also shows the ISM absorption profile obtained from regions A, A+B and C revealing that region C has comparable absorption at high outflow velocities ($|v|\gtrsim250$ \kms) but lower covering fraction at $v\sim0$ relative to the other regions. This suggests that spatial variation in the observed outflow kinematics in this galaxy is due to the paucity of slower moving gas at systemic velocities, i.e., lower covering fraction at $v\sim0$. We quantify this further by considering spatial maps of the covering fraction across the arc. Figure~\ref{fig:coveringfraction-cswa13} shows the spatial map of the best-fit covering fraction profile ($C_f$) at different velocities along with a histogram of its spatial variation. We find that the gas at higher outflow velocities has a relatively uniform covering fraction, with mean $C_f = 0.22$ and 1$\sigma$ spatial scatter of 0.06 at $v = -400$~\kms. At the systemic velocity, the covering fraction varies with mean and spatial scatter $C_f = 0.39\pm0.24$. Specifically, at $v\sim0$, region A has a $C_f\sim0.6$ compared to $C_f\sim0.1$ in region C.

We can summarize the variations in outflow velocity as having a mean and spatial scatter of $v_{out}=\voutValue$~\kms. The covering fraction varies both spatially and spectrally, with typical value and scatter of approximately $C_f=\cfValue$.

\subsection{Column density of \ion{H}{1}}\label{subsec:voigt-fit}
The low-ionization phase of outflowing gas is dominated by \ion{H}{1}, whose column density we can measure directly from \Lya\ absorption. This provides important information on the total outflowing gas mass. 
We use a linear combination of Starburst99 templates \citep{Leitherer1999} and a Voigt profile to simultaneously fit the \Lya\ absorption as well as the stellar continuum \citep{chisholm_SB99, classy7_weida_voigt}. During fitting, we mask out the strong stellar and interstellar features as well as the \Lya\ emission region, and fix the velocity centroid of the \Lya\ absorption to be the same as the $v_{50}$ obtained from ISM absorption lines. We allow the extinction $E(B-V)$, Doppler parameter $b$, column density $N_{HI}$, and covering fraction $f_{\rm cov}$ to be free parameters. The best fit values are obtained from taking the median and Median Absolute Deviation (MAD) from running 100 independent realizations.
Here we refer to covering fraction as $f_{cov}$ in the context of \Lya\ Voigt profile fitting, where $f_{cov}$ can be thought of as the typical value for high column density gas. The covering fraction $C_f(v)$ derived from metal absorption lines is used when we are considering the velocity structure.

Figure~\ref{fig:lya-cswa13} shows the resulting best fit to the spectrum of regions A, A+B and C. In the brightest region A+B of the galaxy, we obtain a value of $\log(N_{HI})=\colDenValue$ for the column density of the \ion{H}{1} gas and a covering fraction of $f_{cov}=\fcovValue$. This gives us a mean column density along the line of sight as $\log(N_{mean}) = \log(N_{HI}\times f_{cov}) = \colDenCovValue$. This is characteristic of damped \Lya\ absorption systems and is comparable to the column densities seen in other star-forming lensed galaxies at similar redshift \citep[e.g.,][]{tucker-dustinthewind}. In contrast, region C does not show such strong \Lya\ damping wings and we find an order of magnitude lower $N_{mean}$ compared to region A+B. We note that the $f_{cov}$ values obtained from this fitting routine are consistent with the $C_f$ obtained from independently fitting the ISM absorption lines. Based on these findings, we use the $N_{mean}$ (Table~\ref{tab:quantities-measured-summary}) obtained from region A+B as the dominant outflow mass component for estimating the mass loss rate and mass loading factor, discussed further in Section~\ref{subsec:masslossrate}.

\subsection{\ion{Si}{2}* emission line map}\label{subsec:emission-linemap}
In this section, we examine fluorescent \ion{Si}{2}* emission to establish the spatial profile of outflowing gas.
\ion{Si}{2}* emission predominantly occurs when a Si$^+$ ion absorbs a photon from the ground state and subsequently decays to an excited fine structure ground state, producing a photon with slightly lower energy (longer wavelength) than the one originally absorbed. These \ion{Si}{2}* transitions appear to be optically thin in our target, such that the fluorescent emission directly traces the spatial distribution of \textit{absorbing} gas \citep[e.g.,][]{jones2012,Prochaska2011}, which is dominated by the outflowing component. 
We use the \ion{Si}{2}*~$\lambda$1533 line (Figure~\ref{fig:spectra-arc}) which is detected across the entire arc and free from intervening absorption. However, it falls within the broad \ion{C}{4} stellar wind feature. Therefore, we fit a small region around \ion{Si}{2}*~$\lambda$1533 with a Gaussian emission line profile combined with a linear continuum to account for the slope of the stellar wind feature.

Figure~\ref{fig:fineemission} (top panel) shows the resulting spatial map of fine structure emission flux in both the image and source plane, with the underlying continuum removed. We find that \ion{Si}{2}* emission is patchy but spatially extends across the entire galaxy. 
The emission is strongest around regions A and B, which have both stronger continuum emission and higher \ion{H}{1} column density than region C (Section~\ref{subsec:voigt-fit}). Together with the velocity measured from corresponding absorption lines, this indicates that the bulk of outflowing mass is associated with regions A and B.

Figure~\ref{fig:fineemission} also shows the continuum and \ion{Si}{2}* spatial profiles extracted from a pseudo-slit through region A+B,  probing the minor axis of the galaxy. The emission is well detected to a radial distance of $R\sim3.5$ kpc. \ion{Si}{2}* is more extended than the stellar continuum but with a rapidly declining flux profile. This suggests that the majority of outflow column density seen in absorption arises from gas confined to small radii (and impact parameters), $R\lesssim5$ kpc, which is supported by observations from galaxy-galaxy pairs and quasar sightlines \citep[e.g.,][]{steidel2010, NielsonMagiicatFitVals,kvgc_ESI2022}.

The spatial extent of outflows as probed in emission by \ion{Si}{2}*, in combination with the kinematics and column density discussed previously, allows a direct measurement of outflow mass loss rate in the low-ionization phase. We discuss the mass loss and its implications in the following section.

\begin{figure*}
   \centering
    \hspace{-1cm}\raisebox{1cm}{\includegraphics[width=0.25\linewidth]{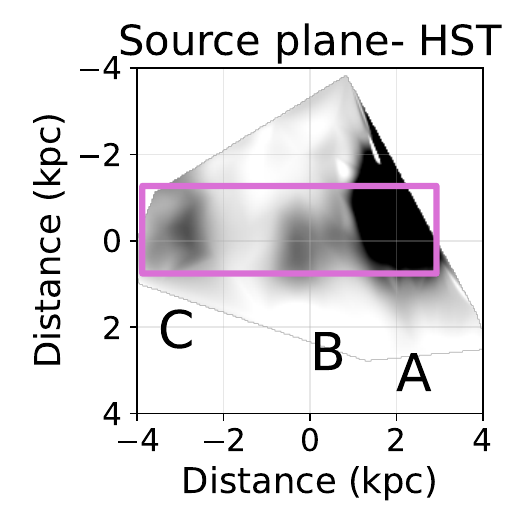}}   \hspace{0cm}
    \includegraphics[width=0.4\linewidth]{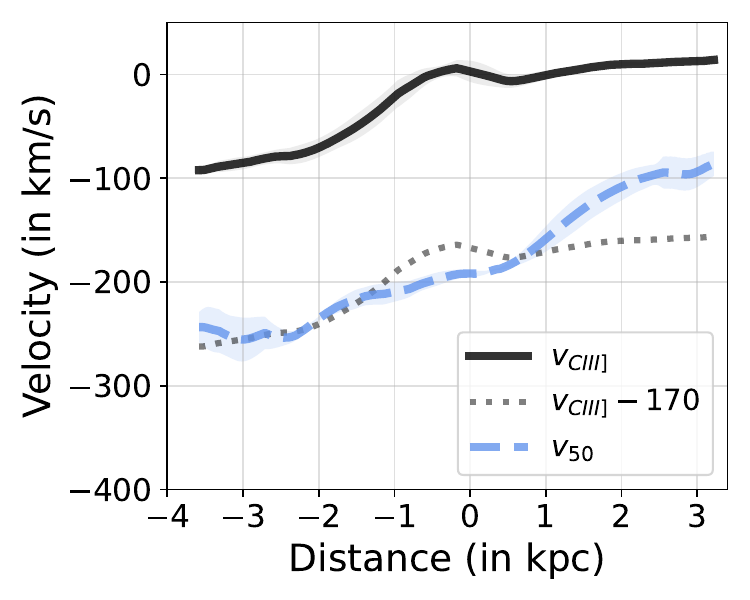}     
    \raisebox{0.5cm}{\includegraphics[width=0.35\linewidth]{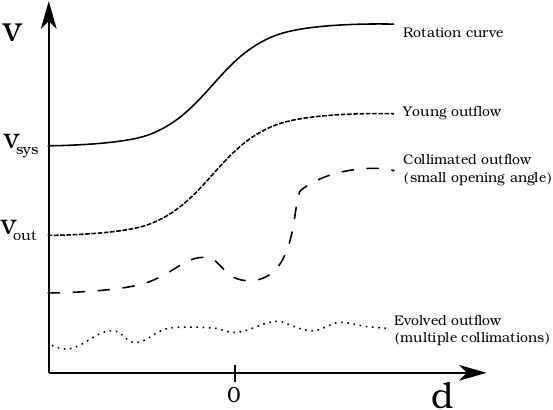}}

    \caption{\emph{Left:} Source plane HST image with a pseudo-slit (purple) along the major axis, used to characterize the outflow velocity structure. 
    \emph{Center:} Systemic and outflow velocity as a function of distance measured along the pseudo-slit. The black and blue lines represent the median systemic and outflow velocities, respectively. The shaded regions show the 1$\sigma$ scatter from collapsing the slit. The gray dashed line shows the $v_{CIII]}$ median line offset by $-170 \kms$, showing an approximately constant outflow velocity relative to the local systemic redshift. 
    {\emph Right:} Schematic illustration of different stages in the evolution of a galactic scale wind (with velocity $v$) as a function of distance from the center of a galaxy ($d$). The rotation curve of the galaxy tracing the motion of the stars is denoted by $v_{sys}$. With respect to this systemic velocity ($v_{sys}$), three different stages of an outflow are shown: (a) young outflow ($v_{out} = v_{sys} + \text{constant}$), (b) collimated outflow ($v_{out} = v_{sys} + \text{constant} + f(r)$), (c) evolved outflow ($v_{out} = \text{constant}$). Each of these scenarios can be distinguished observationally by measuring $v_{out} - v_{sys}$ versus $d$. Based on our schematic, CSWA13 follows the young outflow case, with outflows mirroring the nebular kinematics.
    }
    \label{fig:rotation-outflow-cswa13}
\end{figure*}

\section{Results and Discussion}\label{sec:results-and-discussion}
\subsection{Absorption traces recently-launched outflows}\label{subsec:young-outflows}

Comparing the spatial maps of outflow velocity and nebular emission kinematics allows us to examine whether the outflows are associated with local launching sites, or galaxy-wide winds. 
We consider three example scenarios in the evolution of a galactic-scale wind: (a) young outflow, where the gas is still located close to its launching site; (b) collimated outflow, launched from the central regions of a galaxy with a modest opening angle; and (c) evolved outflow, with coherent motion at distances larger than the galaxy stellar radius.

In the early stages of outflow, feedback processes (e.g., radiation pressure and supernovae) drive the ISM outward from regions of recent star formation. If the initial launching velocity is relatively uniform across different regions in a galaxy, then we would expect the galaxy's rotation curve ($v_{sys}$, defined as the local velocity of stars and \HII regions) to be imprinted in the outflow: $v_{out} \approx v_{sys} + \text{constant}$. We consider this scenario where the ISM is being pressurized by feedback but is yet to burst out of the confines of the galaxy as a 'young outflow.'

As momentum continues to build up in the ISM, the gas can be preferentially launched from regions of low density (e.g., orthogonal to a galactic disk) resulting in a collimated biconical outflow. In this case, the outflow velocity would be higher in the region of collimation compared to the rest of the galaxy, i.e., $v_{out} = v_{sys} + \text{constant} + f(r)$ where $f(r)$ is a function of galactic radius. After a sufficient time, the wind may travel well beyond the galaxy's stellar radius and mix with the CGM, with any signal of the initial launching momentum being mixed such that the outflow velocity appears relatively uniform in down-the-barrel sightlines: $v_{out} = \text{constant}$. As galaxies evolve in time with multiple feedback episodes, we would expect the relation between the systemic and outflow velocity to be a combination of all of these scenarios. 

Figure~\ref{fig:rotation-outflow-cswa13} (right) shows an illustration of these three scenarios. The observed outflow velocity profile of CSWA13 closely follows that of the nebular velocity ($v_{sys}$) with a constant offset, as shown in the top right panel of Figure~\ref{fig:rotation-outflow-cswa13}. 
This is also evident in the spatial maps discussed earlier (Figure~\ref{fig:sourceplane-velmaps}) which show similar gradients. We thus conclude that the outflows seen in absorption are dominated by the 'young outflow' scenario. Similar spatially resolved studies, although at coarser resolution, have found that in a $z=4.9$ arc \citep{swinbank-youngoutflows-2009} the outflows mirror the nebular emission similarly to CSWA13, whereas in the cosmic horseshoe \citep[$z\sim2.4$;][]{bethanjames-horseshoe-muse} the velocity follows the evolved outflow case.

We find a median and $1\sigma$ scatter in velocity centroid of $v_{50} = \voutValue$ \kms\ from spatially resolved regions in CSWA13 (Section~\ref{subsec:measure-v50}). This is similar to down-the-barrel integrated absorption profiles at $z\sim2$ (mean $v_{50}=-141$ and sample scatter $\sim$100 \kms) from \citet{kvgc_ESI2022} and \citet{steidel2010}. However, when we consider the local outflow velocity relative to the star formation regions (i.e., $v_{50} - v_{C III]}$), the scatter is only 41~\kms. This spatial variation in outflow velocity is similar to the $\pm40$~\kms\ differences found in a lensed $z\sim1.7$ galaxy by \cite{boordoloi_rcsga}. 
Thus we find that the outflow velocity is closely connected to the bulk motion of star-forming gas in the galaxy.

\subsubsection{Outflows are encapsulated within the continuum}

Based on our previous discussion, if the outflows are indeed young then we expect a small radial extent. In this case, the fluorescent fine structure emission should be closely connected to the stellar morphology traced by continuum emission. 
We can estimate a characteristic timescale of the ongoing star formation as $\mathrm{sSFR}^{-1} \simeq 25$~Myr \citep{ramesh_CSWA}. A galactic wind launched $\sim$25 Myr ago with constant velocity $-150$~\kms\ would travel a radial distance of $\sim 4$ kpc. In contrast, for an older evolved outflow we would expect fine structure emission to arise at larger radial galactocentric distances. 
From the 2D maps of continuum and fluorescent \ion{Si}{2}* emission (Figure~\ref{fig:fineemission}; described in Section~\ref{subsec:emission-linemap}), the \ion{Si}{2}* is detected across the spatial extent of the galaxy and is patchier. This suggests that the radial extent of the outflowing gas is comparable to the projected size of the galaxy which is $\lesssim8$ kpc.

One might also expect that turbulent young outflows at close radial distances would entrain the ambient gas surrounding the \ion{H}{2} regions \citep[e.g.,][]{mckee_1977} resulting in broad emission lines. Visually inspecting the \ion{C}{3}] and \ion{O}{3}] nebular emission lines in region A+B where the SNR is high, we find that the nebular lines show a clear blueshifted wing component, with the velocity centroid of the blueshifted component similar to that seen in the absorption. This suggests that the outflow emission originates from the same regions as the absorption and supports the idea of a multi-phased wind being launched across the galaxy. This may also enable thorough mixing of the ambient ISM.

\begin{figure*}[!ht]
    \centering
    \vspace{-1cm}
     \includegraphics[width=\linewidth]{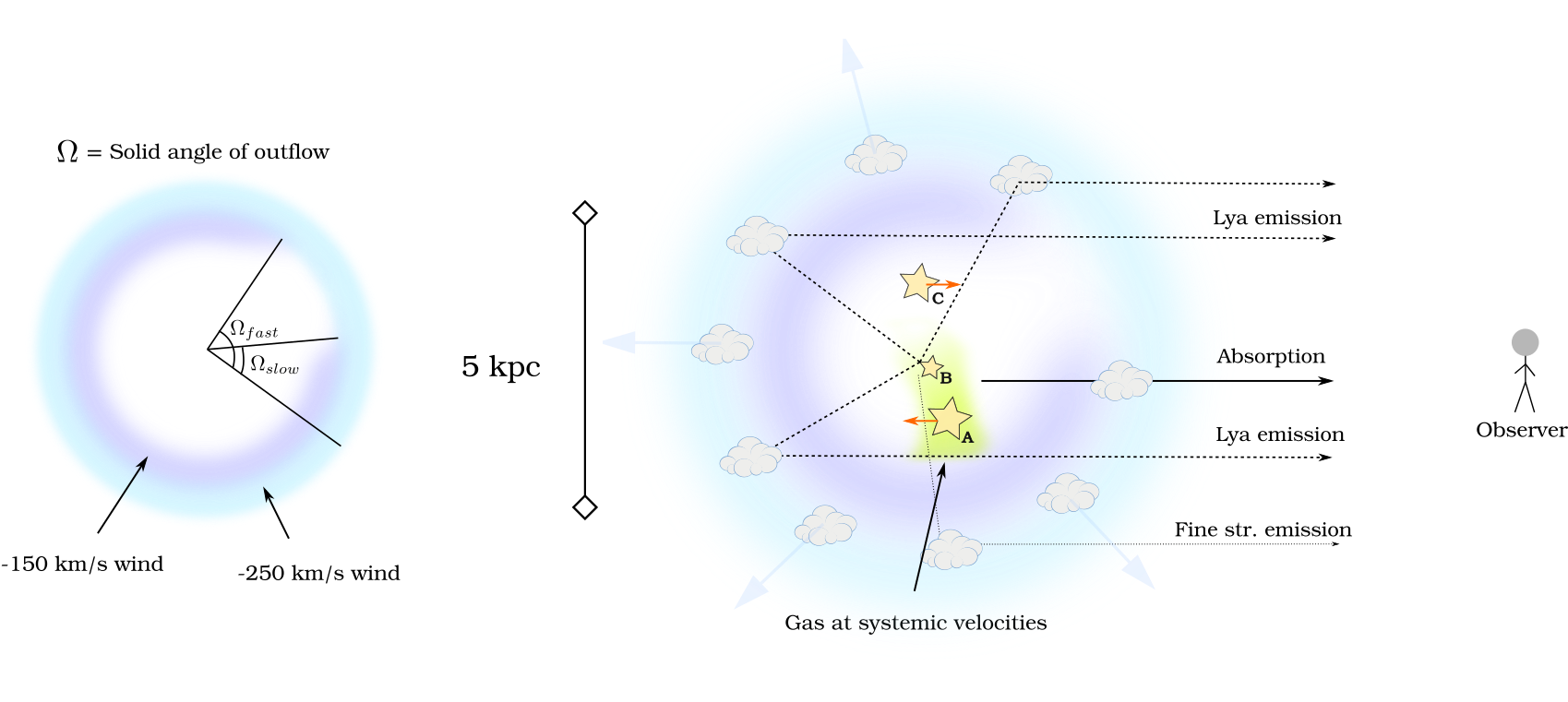}
     
    \caption{Illustration of the observed low-ionization outflow kinematics in CSWA13 using a spherical shell geometry. Gas at systemic velocities ($v\sim0$) is shown in green. The blue and purple regions denote the azimuthal variation of the outflowing ISM gas at $v=-150$ and $v=-250$ \kms, with solid angles $\Omega_{\text{slow}}$ and $\Omega_{\text{fast}}$ respectively. Our observations of the covering fraction of outflowing gas (Figure~\ref{fig:coveringfraction-cswa13}) suggest that gas at higher outflow velocities is more homogenous compared to the slower moving gas. The receding side of the outflow causes backscattering of \Lya\ photons from \ion{H}{1} atoms, resulting in redshifted \Lya\ emission which we detect across the entire galaxy (Figure~\ref{fig:lya-cswa13}). We also observe blueshifted \Lya\ photons leaking along the line-of-sight of region C in the galaxy from forward scattering, which is likely due to low column density of neutral gas at systemic velocities. The orange arrows denote kinematics of the nebular gas (Section~\ref{subsec:neb-kinematics}) in the galaxy.
    }
    \label{fig:illustration-cswa13}
\end{figure*}

\subsection{Outflowing gas is inhomogeneous}\label{subsec:pathy-outflows}

It is clear from the previous subsection that the outflows detected ubiquitously across the galaxy are likely young and radially confined within $\lesssim 8$ kpc. In this subsection, we explore the `geometry' of the outflow as revealed by the variation in covering fraction $C_f(v)$ of the ISM gas at different velocities. $C_f(v)$ is measured from ISM absorption profiles in each spaxel (using double Gaussian fits; Section~\ref{subsec:measure-v50}). Assuming our line-of-sight is representative, the covering fraction is related to the outflow solid angle $\Omega$ as $C_f(v) = \Omega(v)/4\pi$, wherein a spherically symmetric outflow would correspond to $C_f=\text{constant}$ in all spatial regions. 

Figure~\ref{fig:coveringfraction-cswa13} plots the  covering fraction ($C_f$) at $v=-400, -250, -100$ and $0$ \kms\ which correspond roughly to $2.7\times$, $1.7\times$ and  $0.7\times$  the median outflow velocity ($v_{50}$) of the galaxy. We note that the data plotted here is not corrected for the instrumental line spread function. This has the effect of decreasing $C_f$ at the tails of the distribution and increasing it near the $v_{50}$ velocities. Nonetheless, the velocity channels in Figure~\ref{fig:coveringfraction-cswa13} are nearly independent, and the spatial variation in $C_f$ is robust to spectral resolution effects. 
At slower outflow velocities ($v\gtrsim-100$\kms), we can clearly see that the covering fraction of gas in regions A and B of the galaxy is significantly larger by up to $\sim6\times$ than in region C. However, at faster outflow velocities $v\lesssim-250$~\kms, the covering fraction is approximately uniform across the galaxy. This suggests that the high-velocity gas is more homogeneous compared to the slower moving gas. Additionally, we do not find any significant difference between the low-ionization absorption profiles and the intermediate-ionization species (e.g., \ion{Al}{3}, \ion{C}{4})
suggesting that the covering fraction does not vary significantly between these ionization states. We note however that the mean covering fraction measured at different outflow velocities is $\overline{C_f}\sim0.4$ (corresponding to a solid angle of $\Omega=1.4\pi$ steradians if our sightline is representative), indicating that the overall geometry of the outflow is patchy and asymmetric. 

\begin{figure*}
    \centering
    \includegraphics[width=0.49\linewidth]{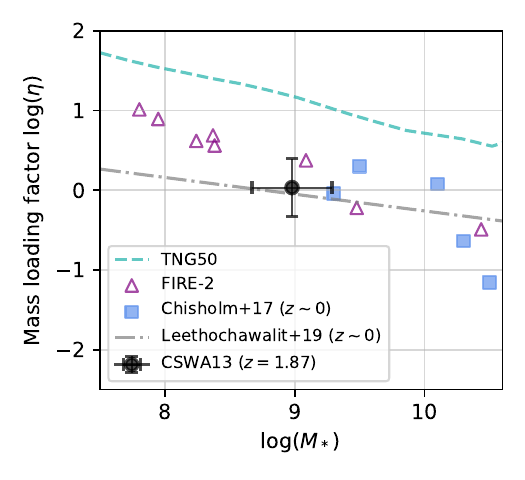}
    \includegraphics[width=0.501\linewidth]{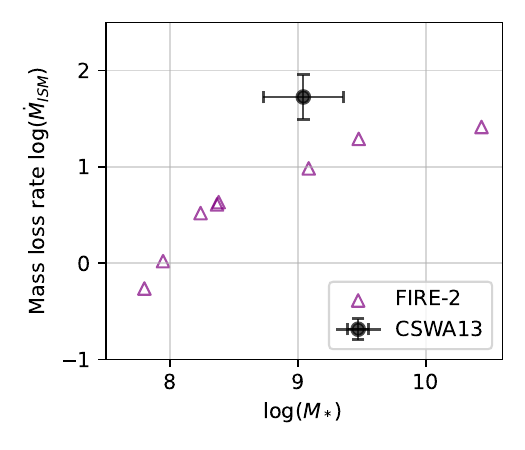}    

    \caption{Comparison of mass loading factor ($\eta = \dot{M}_{out} /$SFR; left) and mass loss rate ($\dot{M}_{out}$; right) obtained for CSWA13 from this work (black point) with cosmological simulations (TNG50, FIRE-2) and galaxies in the low redshift universe. The average $\eta$ obtained from the FIRE-2 \citep{pandya2021} simulations at $z\sim2$ are marked in purple triangles measured at a radial distance of $R=0.1-0.2~R_{vir}$. The green dashed line denotes the median $\eta$ value obtained from simulated $z=2$ galaxies in TNG50 \citep{tng50} at $R=10$ kpc. The $\eta$ measured in local star-forming regions from \citet{chisholm_mass_loading_factor} are shown in blue. The gray dashed line shows the best fit from low-redshift quiescent galaxies \citep{nicha2019}. 
     }
    \label{fig:massloadingfactor}
\end{figure*}

Figure~\ref{fig:illustration-cswa13} illustrates our findings on the spatial and velocity-resolved $C_f(v)$ with a simple spherical shell outflow schematic. 
We find further evidence of the patchiness of outflowing ISM gas from variations in \Lya\ absorption across different regions of the galaxy (e.g., Figure~\ref{fig:lya-cswa13}). 
Region A has damped \Lya\ absorption and exclusively redshifted emission, whereas region C does not show clear damping wings and exhibits a blueshifted emission component. This indicates a significantly lower column density of \HI and associated low-ionization gas toward region C, along with the lower covering fraction (Figure~\ref{fig:coveringfraction-cswa13}).

\subsection{Mass loss rate is comparable to the star formation rate}\label{subsec:masslossrate}

We now use the measurements of outflow geometry, velocity, and column density to estimate the mass loss rate and mass loading factor. Thanks to the spatial and spectral resolution of KCWI data, we can largely avoid systematic uncertainties arising from low spectral resolution, unknown radial distribution of outflowing gas, geometry, and/or outflow velocity relative to systemic, which have presented challenges for earlier efforts \citep[e.g.,][]{pettini_cb58_2000,chisholm_mass_loading_factor}.

If the outflowing gas is described by a spherical shell geometry, with average radial distance $R$ and width $\Delta R$, then we can estimate the mass loss rate through the shell as follows:

\begin{equation}\label{eq:masslossrate}
  \begin{aligned}
\dot{M}_{out} (v_{50})&= (4\pi)(\mu m_p) \times (f_{\rm cov} N_{HI}) \times v_{50} \times R \times \frac{R}{\Delta R} \\  
    & = (10.3\mu) \times  \left(\frac{f_{\rm cov}N_{HI}}{10^{21}\ \text{cm}^{-2}}\right) \times \\
    & \left(\frac{v_{50}}{-10^2\kms}\right) \times
     \left(\frac{R}{1\ \text{kpc}}\right) \times \left(\frac{R}{\Delta R}\right) \Msun \text{yr}^{-1} 
\end{aligned}
\end{equation}
where $\mu\approx1.4$ is the average mass per hydrogen atom in the outflow (mainly accounting for hydrogen and helium), $f_{\rm cov}$ is the mean covering fraction of the outflowing ISM gas over the entire solid angle of $4\pi$ steradians, $N_{HI}$ is the average column density obtained from the integrated spectrum of the galaxy, and $v_{50}$ is the median outflow velocity. The negative sign in velocity indicates that the gas is outflowing from the galaxy. This is similar to the formalism used in \citet{tucker-dustinthewind} and \citet{pettini_cb58_2000}, with the normalization in Equation~\ref{eq:masslossrate} chosen such that each term in the parentheses is $\sim 1$ for our target. Rewriting Equation~\ref{eq:masslossrate} in logarithmic units, with $\mu=1.4$ and $N_{\rm mean} = f_{\rm cov}\times N_{HI}$ (discussed in Section~\ref{subsec:voigt-fit}), we obtain
\begin{equation}\label{eq:masslossrate_log}
  \begin{aligned}
 \log\frac{\dot{M}_{out} (v_{50})}{\Msun \text{yr}^{-1}}  & = 1.159 + \log \left(\frac{N_{\rm mean}}{10^{21}\ \text{cm}^{-2}}\right)  \\
    & + \log\left(\frac{v_{50}}{-10^2\kms}\right) + \log    \left(\frac{R}{1\ \text{kpc}}\right)  \\
    & + \log \left(\frac{R}{\Delta R}\right)  .
\end{aligned}
\end{equation} 
If the ratios in each term are of order unity, then the inferred mass loss rate is $\dot{M}_{out} \sim 10 \, \Msun \text{yr}^{-1}$. 

Throughout this paper (Section~\ref{subsec:measure-v50},~\ref{subsec:voigt-fit},~\ref{subsec:emission-linemap}), we have used the spatially resolved data to constrain each term in Equation~\ref{eq:masslossrate_log}. We summarize these values in Table~\ref{tab:quantities-measured-summary}. The median and $1\sigma$ spatial scatter of the covering fraction of the gas is $\overline{C_f}=\cfValue$ outflowing at $v_{50} = \voutValue~\kms$. 
Based on the fluorescent emission, we find that the wind is confined to a radius similar to that seen in the stellar continuum ($R\sim$\RValue) with the thickness of the shell $\Delta R\sim$\deltaRValue. The mean \ion{H}{1} column density of gas in the outflow is $\sim 10^{21}$~cm$^{-2}$ with region A+B having the dominant contribution. Based on these measurements, we estimate the mass loss rate for the low ionization phase as $\log(\dot{M}_{out} / (\Msun \text{yr}^{-1})) = \mDotValue$.

We note that our measurements of the mass loss rate do not account for the systemic ISM component. However, we obtain consistent results when considering only the outflowing gas. We assess this using the ISM absorption profiles, assuming that they trace the \ion{H}{1} column density. We fit a double Gaussian profile to the absorption profiles of the integrated spectrum and of region A, such that one of the Gaussian components is fixed at $v=0$ and the other component traces only the outflow. We find that while the $v_{50}$ of the outflow component increases, the measured equivalent width that traces the outflow column density decreases by approximately the same factor of $\sim 1.7 \times$. Thus, these effects act in opposing directions such that the measured mass loss rate considering only the blueshifted outflow component would increase by $\lesssim10\%$.

\begin{deluxetable*}{|c|c|c|c|}
    \tablecaption{Summary of measured quantities.}  
    \tablewidth{\textwidth}
    \tabletypesize{\small}
    % Header %
    \tablehead{\colhead{Quantity} & \colhead{}  & \colhead{Measured value} & \colhead{Reference} 
    }
    %data %
    \startdata   
      Stellar Mass &   $\log M_*^{a}$ &  $\StelMassValue$ $\Msun$& Section~\ref{sec:lensmodel}  \\  
      Star Formation Rate &   $\log$ SFR$^{a}$ & $\SFRValue$ $\Msun\  \text{yr}^{-1}$ & Section~\ref{sec:lensmodel}\\
      Outflow velocity Centroid &   $v_{50}$  &  $\voutValue$ \kms & Section~\ref{subsec:measure-v50}, Figure~\ref{fig:sourceplane-velmaps} \\
      Mean Column Density &   $\log(N_{\text{mean}})^{b}$ & $\colDenCovValue$ cm$^{-2}$ &  Section~\ref{subsec:voigt-fit}, Figure~\ref{fig:lya-cswa13}  \\   
      Radial extent of outflowing gas &   $R$ & \RValue & Section~\ref{subsec:emission-linemap}, Figure~\ref{fig:fineemission} \\
      Thickness of radial shell  &    $\Delta R$ & \deltaRValue & Section~\ref{subsec:emission-linemap}, Figure~\ref{fig:fineemission}  \\
      Mass loss rate &   $\log\dot{M}_{out}^{b}$  & $\mDotValue$  $\Msun\ \text{yr}^{-1}$&   Section~\ref{subsec:masslossrate}, Figure~\ref{fig:massloadingfactor}\\         
      Mass loading factor &   $\log\eta^{c}$ & $\etaValue$ & Section~\ref{subsec:masslossrate}, Figure~\ref{fig:massloadingfactor}\\ 
\enddata
\tablenotetext{}{a - Corrected for lensing magnification $|\mu|$.\\ 
b - Here $N_{\text{mean}} = N_{HI}\times f_{cov}$ where $f_{cov}$ is obtained from fitting the \Lya\ profile, as opposed to $C_f(v)$ obtained from the absorption profile of metal ion transitions. \\
c - Assuming a spherical geometry with $\frac{R}{\Delta R}=1.75$} \label{tab:quantities-measured-summary}.
\end{deluxetable*}

The derived mass loss rate is similar to predictions from FIRE-2 simulations at $z\sim2$ measured at a radial distance of $R=0.1-0.2~R_{vir}$ given the stellar mass of CSWA13 (Figure~\ref{fig:massloadingfactor}). We note that this value for the low-ionization outflowing gas represents a lower limit on the total mass loss rate of the galaxy, which likely has contributions from other ionization phases (i.e., ionized and molecular hydrogen). 
For example, analysis of the FIRE simulations by \citet{muratov2017} suggests that $\sim70\%$ of the outflowing and circumgalactic gas is in the low-ionization phase for galaxies with similar mass and redshift as CSWA13. Including the contribution from other phases would therefore increase our estimate of the \textit{total} mass loss rate $\dot{M}_{out,~total}$ by $\sim0.2$ dex. 
The contribution of different ionization states can also be quantified with column densities of various metal ions measured from their absorption lines \citep[e.g.,][]{Chisholm2016,tucker-dustinthewind}, although this is beyond the scope of this paper.

Despite the measured high mass loss rate in CSWA13, the bulk of the ISM gas entrained in the outflow is likely unable to escape its gravitational potential well. We find that the outflow velocity ($v_{50}\simeq-150~\kms$) is lower than the escape velocity estimated from the stellar mass, via the stellar-to-halo mass relation \citep[e.g.,][]{Behroozi19} or estimated escape velocities in FIRE-2 simulated galaxies with similar stellar mass \citep[][]{kvgc_ESI2022}. This suggests that the gas launched via outflows from the ISM will remain bound within the halo and/or recycle back at later times. Given a constant mass loss rate, CSWA13 would need only $\sim$20 Myr to enrich its CGM with a gas mass comparable to its stellar mass. The large outflow rate, if sustained, is thus capable of creating a metal-enriched circumgalactic gas reservoir which can in turn sustain future star formation via recycling.

We now turn to the {\it efficiency} of stellar feedback in driving these powerful outflows. This is quantified by the mass loading factor ($\eta = \frac{\dot{M}_{out}}{\mathrm{SFR}}$), defined as the ratio of the mass loss rate of outflowing gas to the SFR of the galaxy. Cosmological simulations such as TNG50~\citep{tng50} and FIRE-2~\citep{pandya2021} predict that galaxies at $z\sim2$ are highly efficient at driving outflows with typical mass loading factors ranging from $\log\eta\sim0$--1.7 (factors $\eta\sim1$--50) in the stellar mass range $\log M_*/\Msun =8$--10 (Figure~\ref{fig:massloadingfactor}). For CSWA13, using our resolved observations, we measure the mass loading factor for the low-ionization gas phase as $\log\eta = \etaValue$. This efficiency value is similar to predictions from FIRE-2, nearby star-forming galaxies \citep{chisholm_mass_loading_factor}, low-redshift quiescent galaxies \citep{nicha2019} as well as lensed quiescent galaxies at $z\sim1$ \citep{Zhuang2023}, but it is an order of magnitude lower than those predicted by TNG50 at similar redshifts (Figure~\ref{fig:massloadingfactor}). We note that this is not strictly a direct comparison as the methods employed to observationally estimate $\eta$ differ, and the simulation values correspond to a fixed radii (e.g., $R=10$ kpc for TNG50) and thickness (e.g., 0.1 -- 0.2 $R_{vir}$ for FIRE-2) and can make use of full spatial and kinematic information as opposed to down-the-barrel observations of a cylindrical sightline. Nevertheless, our spatially resolved observations serve as an excellent test of different feedback prescriptions, and we view further direct comparison with simulations as a promising prospect. 

\subsection{Spatial variation in outflow properties}\label{subsec:spatial-variation-discussion}

We have demonstrated significant spatial variation in the outflow properties of CSWA13, with the higher surface brightness regions A+B also having stronger outflows (i.e., larger mass loss rates) compared to region C. We also observe variation in the velocity structure, with larger effective outflow velocity in region C. One might expect a correlation of higher star formation densities leading to higher outflow velocities \citep[e.g.,][]{heckman2002, sdss-outflows2016}, in contrast to our results. However, the higher outflow velocity in region C is driven by a lower $C_f$ at low velocities. This may indicate less mass loading of the ambient ISM from region C, resulting in higher velocity from momentum conservation. This is supported by \Lya\ measurements indicating a lower \ion{H}{1} column density and mass loading factor toward region C. 

The complex morphology of CSWA13, aided by gravitational lensing, illustrates the value of spatially resolved information for characterizing gas outflows. We have found order-of-magnitude variation in the total column density toward different regions of the galaxy, with resolved spectroscopy pinpointing regions A+B as the dominant outflow launching sites. We also observe variation in $C_f$ and \Lya\ emission profiles. The lower covering fraction and blueshifted \Lya\ emission component in region C may be particularly interesting in terms of understanding how ionizing photons escape from galaxies, as these signatures are indicative of significant ionizing escape fractions \citep[e.g.,][]{Verhamme2008,jones2013, nicha2016-escape-fraction}. 
The rich variations revealed in this galaxy clearly demonstrate inhomogeneous outflow properties, and the value of spatially resolved information.

\section{Conclusions}
\label{sec:conclusions}

In this paper, we have investigated the spatially resolved outflow properties and kinematics of a $z\sim2$ gravitationally lensed star-forming galaxy (CSWA13) using Keck/KCWI. We map outflows in multiple ultraviolet ISM absorption lines, along with fluorescent \ion{Si}{2}* emission tracing the outflow spatial structure, and nebular emission from \ion{C}{3}] tracing the systemic redshift and velocity structure. We summarize our key findings below.

\begin{enumerate}
    \item The spatial structure of outflow velocity resembles that of the nebular kinematics, which we interpret to be a signature of a young galactic wind that is pressurizing the ISM of the galaxy.
    \item  From the radial extent of \ion{Si}{2}* emission, we estimate that the outflow is largely encapsulated within \RValue. We explore the geometry (e.g., patchiness) of the outflow by measuring the covering fraction at different velocities, finding that the maximum covering fraction is at velocities $v\sim-150$~\kms. We find significant variation in the outflow covering fraction near this peak velocity, with lower but more uniform covering fraction in the higher-velocity gas.
    \item We calculate the mass loss rate and mass loading factors from measurements of the outflow velocity, radius, column density, and covering fraction for the low-ionization outflowing gas in CSWA13.  The mass loss rate ($\log\dot{M}_{out} / (\Msun \text{yr}^{-1}) = \mDotValue$) is comparable to the star formation rate ($\log \text{SFR} / (\Msun \text{yr}^{-1}) = \SFRValue$) resulting in a mass loading factor $\log\eta\sim\etaValue$ in the galaxy, indicating efficient coupling of stellar feedback to drive outflowing mass that is likely to remain in the inner circumgalactic medium or be recycled back into the galaxy. This low-ionization outflow rate is a lower limit on the total mass loss rate of the galaxy, although the low ionization phase is likely the dominant contributor. Based on theoretical predictions, we estimate that the total outflow rate is $\sim0.2$ dex higher with all ionization phases included (Section~\ref{subsec:masslossrate}). We compare our measurement with cosmological simulations, finding that the mass loading factor agrees with predictions from FIRE-2 but is lower by an order of magnitude than those seen in TNG50. 
    \item The outflow properties of CSWA13 exhibit significant spatial variation, with the higher surface brightness regions A+B being the dominant launching site of strong outflows (i.e., larger column density and mass loss rate) compared to the lower surface brightness region C. We also observe variation in the velocity structure, with larger effective outflow velocity in region C. Spatially resolved data aided by gravitational lensing is important for capturing the rich variations in inhomogeneous outflow properties in high-redshift galaxies such as CSWA13.
\end{enumerate}

Overall, these findings support a picture in which outflows observed ubiquitously in early star-forming galaxies such as CSWA13 are responsible for transporting large amounts of mass and metals into the inner circumgalactic medium. This process provides a gas reservoir to sustain star formation at lower redshifts. This work represents early results of our ongoing efforts to spatially resolve outflow (and systemic) kinematics and composition in lensed galaxies at cosmic noon ($z\simeq2-3$), and demonstrates the power of sensitive rest-frame ultraviolet IFS to characterize the effects of feedback on the ISM and CGM at these redshifts. An enlarged sample will help to demonstrate scatter in the population and scaling relations with galaxy mass and other properties. Ultimately the methods used herein represent a path toward establishing the cosmic history of baryon cycling and providing a benchmark for comparison with theoretical models of feedback and galactic outflows. 

\section*{Acknowledgements}
We would like to thank Simon Gazagnes for his help with the Voigt fitting code, and Danielle Berg, Bethan James, and John Chisholm for their insightful discussions.
KVGC, TJ, SR, and KM gratefully acknowledge financial support from the National Science Foundation through grant AST-2108515, the Gordon and Betty Moore Foundation through Grant GBMF8549, NASA through grant HST-GO-16773, and from a Dean’s Faculty Fellowship. 
AJS was supported by NASA through the NASA Hubble Fellowship grant HST-HF2-51492 from the Space Telescope Science Institute, which is operated by the Association of Universities for Research in Astronomy, Inc., for NASA, under contract NAS5-26555. This research was supported by the Australian Research Council Centre of Excellence for All Sky Astro-
physics in 3 Dimensions (ASTRO 3D), through project number CE170100013.
The data presented herein were obtained at the W. M. Keck Observatory, which is operated as a scientific partnership among the California Institute of Technology, the University of California and the National Aeronautics and Space Administration. The Observatory was made possible by the generous financial support of the W. M. Keck Foundation. The authors wish to recognize and acknowledge the very significant cultural role and reverence that the summit of Maunakea has always had within the indigenous Hawaiian community. We are most fortunate to have the opportunity to conduct observations from this mountain.

\bibliography{ism_kinematics}

\end{document}